\documentclass[aps, prd, showkeys, singlecolumn, nofootinbib, floatfix, superscriptaddress]{revtex4-2}
\usepackage{amssymb}
\usepackage{amsmath}
\usepackage{graphicx}
\usepackage{dcolumn}
\usepackage{bm}
\usepackage[frak=esstix]{mathalpha}
\usepackage{soul, xpatch, caption, subcaption}
\usepackage[normalem]{ulem}
\usepackage[dvipsnames]{xcolor}
\usepackage[colorlinks=true,allcolors = blue]{hyperref}
\usepackage[multiple]{footmisc}

\begin{document}

\title{Method of moments for a relativistic single-component gas}

\author{Caio V.~P.~de Brito}
\email{caio\_brito@id.uff.br}
\affiliation{Instituto de F\'{\i}sica, Universidade Federal Fluminense \\ Av.~Gal.~Milton Tavares de Souza, S/N, 24210-346, Gragoatá, Niter\'{o}i, Rio de Janeiro, Brazil}
\affiliation{Institute for Theoretical Physics, Goethe University, Max-von-Laue-Str.~1, D-60438 Frankfurt am Main,  Germany}
\author{Gabriel S.~Denicol}
\email{gsdenicol@id.uff.br}
\affiliation{Instituto de F\'{\i}sica, Universidade Federal Fluminense \\ Av.~Gal.~Milton Tavares de Souza, S/N, 24210-346, Gragoatá, Niter\'{o}i, Rio de Janeiro, Brazil}

\begin{abstract}
We derive the equations of motion for all the irreducible moments of the single-particle distribution function. We find that these moment equations of motion are highly coupled, with the dynamics of lower-rank moments always being coupled to those of a higher-rank, leading to an endless tower of equations. Considering a massless gas in Bjorken flow, we investigate how this hierarchy of differential equations can be properly truncated and solved.
\end{abstract}

\maketitle

\section{Introduction}

For the past decades, relativistic fluid dynamics has been successfully used in the description of cold atomic gases \cite{Schafer:2009dj, Kulkarni2012} as well as of the quark-gluon plasma created in ultra-relativistic heavy-ion collisions \cite{gale2013hydrodynamic, Jaiswal_2016, Florkowski:2017ixx}. In order to obtain an accurate description of these systems, it is indispensable to take into account dissipative effects in the differential equations that govern the relativistic fluid-dynamical evolution. This has motivated a general discussion on the derivation of relativistic dissipative fluid dynamics from an underlying microscopic theory \cite{Romatschke:2017ejr, Florkowski:2017ixx, denicolbook} and, furthermore, in establishing the domain of applicability of such fluid-dynamical formulations. In particular, this issue has been widely investigated for relativistic dilute gases using the Boltzmann equation as a starting point \cite{denicolbook}. 

Chapman-Enskog theory \cite{chapman1970mathematical,cercignani,degroot} is the most widespread method to derive fluid dynamics from the Boltzmann equation. In this approach, a particular solution for the single-particle distribution function is found in the form of a gradient expansion. Different truncations of this expansion lead to distinct fluid-dynamical theories: a truncation at zeroth order leads to the Euler equations (ideal fluid dynamics), while at first order it yields the Navier-Stokes equations \cite{landau}. Unlike its non-relativistic counterpart, the relativistic Navier-Stokes theory is ill-defined, as its acausal nature renders the global equilibrium state unstable \cite{pichon:65etude, hw1985}. Moreover, higher-order truncations of the Chapman-Enskog expansion lead to Burnett and super-Burnett theories, which suffer from the so-called Bobylev instability even in the non-relativistic regime \cite{bobylev}. These problems lead to the conclusion that the traditional Chapman-Enskog method does not yield fluid-dynamical formulations that can actually be employed for practical purposes.\footnote{In Ref.~\cite{Denicol:2016bjh}, a possibly \textit{convergent} generalization of the Chapman-Enskog expansion was proposed. However, its convergence has only been investigated in Bjorken flow \cite{Bjorken1983} within the relaxation time approximation \cite{oldRTA}.}

An alternative to obtain fluid-dynamical theories from the Boltzmann equation that does not display non-physical features is the method of moments. It was originally developed by Grad for non-relativistic gases \cite{grad1}, introducing an expansion for the single-particle distribution function in terms of a complete and orthogonal basis of generalized Hermite polynomials \cite{grad2}. Three decades later, Israel and Stewart were the pioneers in extending Grad's work to relativistic systems \cite{iskinetic}. In the absence of a convenient orthogonal basis, they simply expanded the single-particle distribution function using a basis of 4-momenta, $1, k_\mu, k_\mu k_\nu, \cdots$. This expansion is then truncated at second order, so that the distribution function is described solely in terms of degrees of freedom that can be matched to the independent fields appearing in the conserved currents. The truncated distribution function is finally inserted into the first three moments of the Boltzmann equation, leading to the so-called Israel-Stewart transient theory of relativistic fluid dynamics. The derived equations are shown to be linearly causal and stable as long as the transport coefficients satisfy certain constraints, cf.~Refs.~\cite{olson, pushi, bdcoupling, Sammet2023}. We remark that a major difference between the method of moments and the Chapman-Enskog method is the fact that the first relies on a truncation in degrees of freedom, while the latter corresponds to a truncation in a small parameter, characterized by gradients of the fluid-dynamical variables.

In Ref.~\cite{dnmr}, the relativistic version of the method of moments was generalized and the single-particle distribution function was expanded in terms of a basis of \textit{irreducible}\footnote{with respect to Lorentz transformations that leave the fluid 4-velocity unchanged, i.e., $\Lambda^\mu{}_\nu u^\nu = u^\mu$ \cite{degroot}.} momenta, $1, k_{\langle \mu \rangle}, k_{\langle \mu} k_{\nu \rangle}, \cdots$. In contrast to the basis chosen by Israel and Stewart, the irreducible momenta form a complete and \textit{orthogonal} basis \cite{degroot, denicolbook}. Hence, the expansion coefficients can be obtained in exact form without relying on a matching procedure, and their equations of motion stem from the Boltzmann equation. In practice, one replaces the Boltzmann equation by a set of coupled partial differential equations for the moments of the single-particle distribution function, and fluid-dynamical theories are then obtained from a systematic truncation of these equations. As a matter of fact, the method of moments, as proposed in Ref.~\cite{dnmr}, has been consistently employed to obtain causal fluid-dynamical theories from the Boltzmann equation, cf.~Refs.~\cite{Denicol_2012, Denicol:2012yr, Denicol:2019iyh, Rocha:2021lze, Weickgenannt:2022zxs, Weickgenannt:2022qvh, Fotakis:2022usk, Wagner:2022ayd, deBrito:2023tgb, Rocha:2023hts, Chattopadhyay:2023hpd, deBrito:2023tgb}. 

Nevertheless, the convergence of the moment equations has never been fully explored in the relativistic regime. For instance, the equations of motion for the moments are highly coupled, with the dynamics of moments that are of a lower order in the moment expansion coupling to those that are of a higher order. How this hierarchy of equations can be properly truncated and how this truncation quantitatively affects the solutions for each moment is not well known. One reason for this lack of studies is that the equations of motion for the irreducible moments of the non-equilibrium distribution function have only been obtained for a handful of moments \cite{dnmr, deBrito:2023tgb}.\footnote{In Ref.~\cite{Tinti_2019}, general equations of motion for the \textit{reducible} moments of the single-particle distribution function were thoroughly investigated for a system of electrically charged particles, in the context of the Boltzmann-Vlasov-Maxwell equation.}\footnote{During the writing of this manuscript, general moment equations were also derived in Ref.~\cite{ye2023linear}.} 

In the present work, we bridge this gap by deriving the equations of motion for all the \textit{irreducible} moments of the distribution function. Furthermore, we show how these equations simplify in the highly symmetric configuration of Bjorken flow \cite{Bjorken1983}. In this case, we recover the results first obtained in Ref.~\cite{denicolbook}, where the method of moments was developed employing the aforementioned symmetries from the start. This shows the consistency of the derivations developed in the present work. We then investigate the convergence properties of the solutions for this hierarchy of equations for a gas of classical massless particles in Bjorken flow within the relaxation time approximation \cite{oldRTA}. 

This work is organized as follows. In Sec.~\ref{sec:mm} we outline the method of moments and, in Sec.~\ref{sec:eoms}, systematically calculate a hierarchy of equations of motion for the irreducible moments of the distribution function. Then, in Sec.~\ref{sec:bjorken}, we derive the moment equations for a system undergoing Bjorken flow and, in Sec.~\ref{sec:convergence},
we thoroughly analyze the convergence of the solutions for these equations for a classical massless gas undergoing a Bjorken expansion. All our conclusions are summarized in Sec.~\ref{sec:conc}. The details of the derivation of the main results of this paper is delegated to the appendices. Throughout this work, we make use of natural units, $c=\hbar=k_B=1$, and adopt the mostly minus convention for the Minkowski metric tensor, $g_{\mu\nu}=\mathrm{diag} \, (1,-1,-1,-1)$, unless stated otherwise. 

\section{Relativistic Boltzmann equation and the method of moments}
\label{sec:mm}

The relativistic Boltzmann equation is an integro-differential equation that describes the dynamics of the single-particle momentum distribution of a dilute gas \cite{cercignani}
\begin{equation}
k^\mu \partial_\mu f_{\mathbf{k}} = C[f], \label{boltz}
\end{equation}
where $k^\mu = (\sqrt{m^2+\mathbf{k}^2}, \mathbf{k})$ is the 4-momentum, with $m$ being the mass of the particles, $f_{\mathbf{k}}$ is the single-particle distribution function, and $C[f]$ is the collision term. Considering classical particles and taking into account exclusively binary elastic collisions, it reads
\begin{equation}
C[f] = \frac{g}{2} \int dK' dP dP' \mathcal{W}_{\mathbf{k}\mathbf{k}' \leftrightarrow \mathbf{p}\mathbf{p}'}\left( f_{\mathbf{p}}f_{\mathbf{p}'} - f_{\mathbf{k}}f_{\mathbf{k}'} \right).
\end{equation}
Here, $g$ is the degeneracy factor (for simplicity, hereon we assume $g=1$), $dK=d^3 \mathbf{k}/\left[(2\pi)^3 k^0\right]$ is the Lorentz-invariant volume element in momentum space and $\mathcal{W}_{\mathbf{k}\mathbf{k}' \leftrightarrow \mathbf{p}\mathbf{p}'}$ is the (also Lorentz-invariant) transition rate.

It is convenient to decompose the single-particle distribution function into an equilibrium and a non-equilibrium contribution,
\begin{equation}
f_{\mathbf{k}} = f_{0\mathbf{k}} + \delta f_{\mathbf{k}} = f_{0\mathbf{k}} \left( 1 + \phi_{\mathbf{k}} \right ), \label{eq:expand_f}
\end{equation}
where $f_{0\mathbf{k}} = \exp{\left( \alpha - \beta E_{\mathbf{k}} \right)}$ is the usual Maxwell-Boltzmann equilibrium distribution function \cite{reichl1999modern}, with $\alpha = \mu/T$ (referred to as the thermal potential, where $\mu$ is the chemical potential and $T$ is the temperature), $\beta = 1/T$, $E_{\mathbf{k}} = u_\mu k^\mu$ being the energy of the particle in the local rest frame of the fluid, and $u^\mu$ the fluid 4-velocity -- a normalized time-like 4-vector, $u_\mu u^\mu = 1$. Furthermore, $\delta f_{\mathbf{k}}$ denotes the deviation from local equilibrium of the distribution function, with $\phi_{\mathbf{k}} \equiv \delta f_{\mathbf{k}} / f_{0\mathbf{k}}$. In the method of moments, the non-equilibrium distribution function is expanded using a complete and orthogonal basis of irreducible momenta \cite{dnmr},
\begin{equation}
\phi_{\mathbf{k}} = \sum_{\ell = 0}^{\infty} \lambda^{\langle \mu_1 \cdots \mu_\ell \rangle}_{\mathbf{k}} k_{\langle \mu_1} \cdots k_{\mu_\ell \rangle}, \label{eq:expan_phi}
\end{equation}
where $A^{\langle \mu_1 \cdots \mu_\ell \rangle} = \Delta^{\mu_1 \cdots \mu_\ell}_{\nu_1 \cdots \nu_\ell}A^{\nu_1 \cdots \nu_\ell}$ denotes the irreducible projection of an arbitrary tensor $A^{\nu_1 \cdots \nu_\ell}$, with 
\begin{equation}
\Delta^{\mu_1\cdots\mu_\ell}_{\nu_1\cdots\nu_\ell}=\sum_{q=0}^{\left[\ell/2\right]}\frac{C(\ell,q)}{\mathcal{N}_{\ell,q}}\sum_{\mathcal{P}^\ell_\mu\mathcal{P}^\ell_\nu}\Delta^{\mu_1\mu_2}\cdots\Delta^{\mu_{2q-1}\mu_{2q}}\Delta_{\nu_1\nu_2}\cdots\Delta_{\nu_{2q-1}\nu_{2q}}\Delta^{\mu_{2q+1}}_{\nu_{2q+1}}\cdots\Delta^{\mu_\ell}_{\nu_\ell} ,\label{projector}
\end{equation}
being the $2\ell$-index projection operator symmetric under the exchange of $\mu$- and $\nu$-type indices, traceless and orthogonal to the 4-velocity in every index \cite{degroot, denicolbook}. It is constructed in terms of the $2$-index projection operator onto the 3-space orthogonal to the fluid 4-velocity, $\Delta^{\mu\nu} = g^{\mu\nu} - u^\mu u^\nu$. The first sum runs up to the largest integer less than or equal to $\ell/2$, while the second sum accounts for all possible permutations of the indices. The factors $C(\ell,q)$ and $\mathcal{N}_{\ell,q}$ are defined as
\begin{equation}
C(\ell,q)=(-1)^q\frac{(\ell!)^2}{(2\ell)!}\frac{(2\ell-2q)!}{q!(\ell-q)!(\ell-2q)!}, \hspace{.5cm} \mathcal{N}_{\ell,q}=\frac{1}{(\ell-2q)!}\left(\frac{\ell!}{2^q q!}\right)^2. \label{def_CN}
\end{equation}
The former is essential to ensure the traceless property of the projection operator,
\begin{equation}
\Delta^{\mu_1\cdots\mu_\ell}_{\nu_1\cdots\nu_\ell} g^{\nu_i \nu_j}=0, \hspace{.5cm} \Delta^{\mu_1\cdots\mu_\ell}_{\nu_1\cdots\nu_\ell} g_{\mu_i \mu_j}=0, \hspace{.5cm} 1 \leq (i, j) \leq \ell, \label{traceless}
\end{equation}
while the latter is simply the inverse of the total number of permutations given in the second sum in Eq.~\eqref{projector}, in order to avoid over-counting any particular term. 

We remark that, in contrast to the basis chosen by Israel and Stewart, the irreducible momenta satisfy the following orthogonality condition \cite{degroot, denicolbook}
\begin{equation}
\label{eq:ortho_momenta}
\int dK F(E_{\mathbf{k}}) k^{\langle\mu_1}\cdots k^{\mu_\ell\rangle}k_{\langle\nu_1}\cdots k_{\nu_m\rangle}=\frac{\ell! \, \delta_{\ell m}}{(2\ell+1)!!}\Delta^{\mu_1\cdots\mu_\ell}_{\nu_1\cdots\nu_m}\int dK F(E_{\mathbf{k}}) \left( \Delta^{\alpha\beta} k_\alpha k_\beta \right)^m,
\end{equation}
with $F(E_{\mathbf{k}})$ being an arbitrary function of $E_{\mathbf{k}}$. 

Following Ref.~\cite{dnmr}, the expansion coefficients $\lambda^{\langle \mu_1 \cdots \mu_\ell \rangle}_{\mathbf{k}}$ are functions of $E_{\mathbf{k}}$ and are further expanded using a complete basis of orthogonal functions, $P_{\mathbf{k}n}^{(\ell)}$,
\begin{equation}
\lambda^{\langle \mu_1 \cdots \mu_\ell \rangle}_{\mathbf{k}} = \sum^{\infty}_{n = 0} \Phi^{\langle \mu_1 \cdots \mu_\ell \rangle}_n P_{\mathbf{k} n}^{(\ell)}. \label{eq:exp-lambda}
\end{equation}
For the sake of convenience\footnote{For massless particles, they reduce to the associated Laguerre polynomials \cite{denicolbook}.}, $P_{\mathbf{k} n}^{(\ell)}$ are taken as polynomials of $E_{\mathbf{k}}$, 
\begin{equation}
P_{\mathbf{k} n}^{(\ell)} =  \sum^{n}_{r = 0} a_{n r}^{(\ell)} E_{\mathbf{k}}^r.
\end{equation}
Without loss of generality, we set $P_{\mathbf{k} 0}^{(\ell)}=1$, so that $a_{00}^{(\ell)}=1$ and all remaining coefficients can be obtained using the Gram-Schmidt orthogonalization procedure \cite{dnmr}. In particular, these functions are constructed to satisfy the following orthogonality condition
\begin{equation}
\frac{\mathcal{N}^{(\ell)}}{(2\ell+1)!!} \int dK \left(\Delta_{\alpha\beta}k^\alpha k^\beta\right)^\ell  P^{(\ell)}_{\mathbf{k}n}P^{(\ell)}_{\mathbf{k}m} f_{0\mathbf{k}} = \delta_{mn},  \label{eq:ortho-func-P}
\end{equation}
where $\mathcal{N}^{(\ell)} = (-1)^\ell/I_{2\ell, \ell}$ -- see Appendix E of Ref.~\cite{dnmr} or Ref.~\cite{denicolbook} for a detailed derivation -- and $I_{ij}$ are thermodynamic integrals defined as
\begin{equation}
I_{ij} = \frac{(-1)^j}{(2j+1)!!}\int dK E^{i-2j}_{\mathbf{k}} \left(\Delta_{\alpha\beta}k^\alpha k^\beta\right)^j f_{0\mathbf{k}}.
\end{equation}

From the orthogonality condition \eqref{eq:ortho_momenta}, and using Eqs.~\eqref{eq:expan_phi}-\eqref{eq:ortho-func-P}, the expansion coefficients in Eq.~\eqref{eq:exp-lambda} can be expressed as
\begin{equation}
\Phi_n^{\langle \mu_1 \cdots \mu_\ell \rangle} = \frac{\mathcal{N}^{(\ell)}}{\ell!} \sum_{r = 0}^n a_{n r}^{(\ell)} \rho^{\mu_1 \cdots \mu_\ell}_r,
\end{equation}
with $\rho^{\mu_1 \cdots \mu_\ell}_r$ being the irreducible moments of the non-equilibrium distribution function,
\begin{equation}
\rho^{\mu_1\cdots\mu_\ell}_r=\int dK E^r_{\mathbf{k}} k^{\langle\mu_1}\cdots k^{\mu_\ell\rangle} \delta f_{\mathbf{k}}. 
\end{equation}
Finally, we have
\begin{equation}
f_{\mathbf{k}} = f_{0 \mathbf{k}} \left(  1 + \sum_{\ell = 0}^\infty \sum_{n = 0}^{\infty} \sum_{r=0}^{n} \frac{\mathcal{N}^{(\ell)}}{\ell!} a^{(\ell)}_{n r} \, P_{\mathbf{k} n}^{(\ell)} \, \rho^{\mu_1 \cdots \mu_\ell}_r \, k_{\langle \mu_1} \cdots k_{\mu_\ell \rangle} \right ). \label{eq:usual_moments}
\end{equation}

We have expressed the single-particle distribution function completely in terms of the irreducible moments of $\delta f_{\mathbf{k}}$, which are the only terms \textit{a priori} unknown in the expression above. In principle, the exact solution for $f_{\mathbf{k}}$ is obtained when all sums are taken to infinity, although this is unfeasible in practice. Nonetheless, as more terms are included in the expansion, one expects to obtain values of the distribution function which are in better agreement with the \textit{exact} solution of the Boltzmann equation, at least up to a given energy scale. It is then necessary to determine the time evolution of the irreducible moments in order to obtain the single-particle distribution function itself, thus effectively solving the Boltzmann equation. In general, this task requires the inclusion of moments of arbitrarily high ranks and not just those that appear in fluid-dynamical theories. 

So far, equations of motion for the irreducible moments have always been calculated separately (rank by rank), and, to this day, only up to rank 4 \cite{dnmr, deBrito:2023tgb}. Nevertheless, this task becomes progressively more exhausting as one includes moments of higher ranks in the expansion given by Eq.~\eqref{eq:usual_moments}. The main goal of this work is to calculate a general equation of motion for an irreducible moment of arbitrary rank $\ell$, from which the dynamics of \textit{any} irreducible moment can be straightforwardly obtained. 

\section{Equations of motion}
\label{sec:eoms}

First, it is convenient to define the irreducible moments of a \textit{generic} single-particle distribution function, $f_{\mathbf{k}}$, 
\begin{equation}
\varrho^{\mu_1\cdots\mu_\ell}_r=\int dK E^r_{\mathbf{k}} k^{\langle\mu_1}\cdots k^{\mu_\ell\rangle} f_{\mathbf{k}}. \label{eq:def_varvarrho}
\end{equation}
We emphasize that the irreducible moments $\varrho^{\mu_1\cdots\mu_\ell}_r$ are defined slightly different than $\rho^{\mu_1\cdots\mu_\ell}_r$. They are integrals of the single-particle distribution function, $f_{\mathbf{k}}$, rather than integrals of its non-equilibrium component, $\delta f_{\mathbf{k}}$, cf.~Eq.~\eqref{eq:usual_moments}. The latter prescription can be straightforwardly recovered by simply factorizing $f_{\mathbf{k}}$ as in Eq.~\eqref{eq:expand_f}. Nevertheless, we remark that
\begin{equation}
\int dK E^r_{\mathbf{k}} k^{\langle\mu_1} \cdots k^{\mu_\ell\rangle} f_{0\mathbf{k}} = 0, \quad \forall \quad \ell > 0, \label{eq:vanishing_integrals}
\end{equation}
since this is a $\ell$-th rank irreducible tensor that depends solely on the temperature, $T$, chemical potential, $\mu$, fluid 4-velocity, $u^\mu$, and metric tensor, $g^{\mu\nu}$. Meanwhile, it must also be traceless, symmetric and orthogonal to the 4-velocity in every index. However, such tensors cannot be constructed using only $u^\mu$ and $g^{\mu\nu}$, and therefore the above integrals are identically zero, as long as $\ell > 0$. In this context, the definition adopted in Eq.~\eqref{eq:def_varvarrho} will turn out to be a rather convenient choice.

The Boltzmann equation, Eq.~\eqref{boltz}, can be expressed as
\begin{equation}
\frac{d}{d\tau}f_{\mathbf{k}} = \frac{C[f]}{E_{\mathbf{k}}} - \frac{1}{E_{\mathbf{k}}} k_{\langle \mu \rangle} \nabla^\mu f_{\mathbf{k}}, \label{eq:eom-boltz}
\end{equation}
where $d/d\tau = u^\mu \partial_\mu$ is the comoving time derivative and $\nabla^\mu = \Delta^{\mu\nu}\partial_\nu$ is the 4-gradient operator. Using this result, it is possible to show that the irreducible moments of \textit{arbitrary} rank, $\varrho^{\mu_1\cdots\mu_\ell}_r$, satisfy the following equations of motion 
\begin{eqnarray}
\dot{\varrho}^{\langle\mu_1\cdots\mu_\ell\rangle}_r&=&\mathcal{C}^{\mu_1\cdots\mu_\ell}_{r-1}+r\varrho^{\mu_1\cdots\mu_{\ell+1}}_{r-1}\dot{u}_{\mu_{\ell+1}}-\Delta^{\mu_1\cdots\mu_\ell}_{\nu_1\cdots\nu_\ell}\nabla_{\nu_{\ell+1}}\varrho^{\nu_1\cdots\nu_{\ell+1}}_{r-1}+(r-1)\varrho^{\mu_1\cdots\mu_{\ell+2}}_{r-2}\sigma_{\mu_{\ell+1}\mu_{\ell+2}}+\ell\varrho^{\alpha\langle\mu_1\cdots\mu_{\ell-1}}_r\omega^{\mu_\ell\rangle}{}_{\alpha}+\notag \\
&+&\frac{\ell}{2\ell+1}\left[rm^2\varrho_{r-1}^{\langle\mu_1\cdots\mu_{\ell-1}}-(r+2\ell+1)\varrho_{r+1}^{\langle\mu_1\cdots\mu_{\ell-1}}\right]\dot{u}^{\mu_\ell\rangle}+\frac{1}{3}\left[(r-1)m^2\varrho_{r-2}^{\mu_1\cdots\mu_\ell}-(r+\ell+2)\varrho_{r}^{\mu_1\cdots\mu_\ell}\right]\theta+\notag\\
&+&\frac{\ell}{2\ell+3}\left[(2r-2)m^2\varrho_{r-2}^{\alpha\langle\mu_1\cdots\mu_{\ell-1}}-(2r+2\ell+1)\varrho_{r}^{\alpha\langle\mu_1\cdots\mu_{\ell-1}}\right]\sigma_\alpha^{\mu_\ell\rangle}-\frac{\ell}{2\ell+1}\nabla^{\langle\mu_1}\left(m^2\varrho_{r-1}^{\mu_2\cdots\mu_{\ell}\rangle}-\varrho_{r+1}^{\mu_2\cdots\mu_{\ell}\rangle}\right)+\notag\\
&+&\frac{\ell(\ell-1)}{4\ell^2-1}\left[(r-1)m^4\varrho_{r-2}^{\langle\mu_1\cdots\mu_{\ell-2}}-(2r+2\ell-1)m^2\varrho_{r}^{\langle\mu_1\cdots\mu_{\ell-2}}+(r+2\ell)\varrho_{r+2}^{\langle\mu_1\cdots\mu_{\ell-2}}\right]\sigma^{\mu_{\ell-1}\mu_\ell\rangle}, \label{eq:eom_moms}
\end{eqnarray}
where $\theta = \partial_\mu u^\mu$ is the expansion rate, $\sigma^{\mu\nu} = \nabla^{\langle\mu}u^{\nu\rangle}$ is the shear tensor, $\omega^{\mu\nu} = (\nabla^\mu u^\nu - \nabla^\nu u^\mu)/2$ is the vorticity tensor and $\mathcal{C}^{\mu_1\cdots\mu_\ell}_{r}$ is the generalized collision term, defined according to Ref.~\cite{dnmr},
\begin{equation}
\mathcal{C}^{\mu_1\cdots\mu_\ell}_{r} = \int dK E_{\mathbf{k}}^{r} k^{\langle\mu_1}\cdots k^{\mu_\ell\rangle} C[f]. \label{generalized_collision}
\end{equation}
In the derivation of Eq.~\eqref{eq:eom_moms}, we have used the following identities,
\begin{subequations}
\label{eq:rel-irred-moment}
\begin{align}
k^{\langle\mu_1}\cdots k^{\mu_\ell\rangle} &= k^{\langle\mu_1\rangle}\cdots k^{\langle\mu_\ell\rangle} + \sum_{q=1}^{[\ell/2]} \left( \Delta_{\alpha\beta} k^\alpha k^\beta \right)^q \frac{\ell!}{2^q q!}\frac{C(\ell,q)}{\mathcal{N}_{\ell,q}}\sum_{\mathcal{P}^\ell_\mu}\Delta^{\mu_1\mu_2}\cdots\Delta^{\mu_{2q-1}\mu_{2q}}k^{\langle\mu_{2q+1}\rangle}\cdots k^{\langle\mu_\ell\rangle}, \label{identity0} \\
k^{\langle\nu_1}\cdots k^{\nu_\ell\rangle}k^{\langle\nu_{\ell+1}\rangle} &= k^{\langle\nu_1}\cdots k^{\nu_{\ell+1}\rangle} + \frac{\ell}{2\ell + 1} \left( \Delta_{\lambda \beta} k^\lambda k^\beta \right) \Delta^{\nu_1\cdots\nu_\ell}_{\alpha_1\cdots\alpha_\ell}\Delta^{\alpha_\ell\nu_{\ell+1}}k^{\langle\alpha_1}\cdots k^{\alpha_{\ell-1}\rangle}.
\label{identity2}
\end{align}
\end{subequations}
The calculations to obtain Eqs.~\eqref{eq:eom_moms} and \eqref{eq:rel-irred-moment} are reported in detail in Appendices \ref{app_calculations} and \ref{app_formulas}. 

The result obtained in Eq.~\eqref{eq:eom_moms} is consistent with previous calculations for $\ell=0, 1, 2$ \cite{dnmr}, as well as for $\ell=3, 4$ \cite{deBrito:2023tgb}, originally derived in the context of second- and third-order theories of relativistic fluid dynamics, respectively. In particular, we remark that this result is the relativistic generalization of Eq.~(8) from Ref.~\cite{struchtrup2004stable}. Furthermore, we note that Eqs.~\eqref{eq:eom_moms} do not clearly display the traditional Navier-Stokes terms, i.e., terms that are of first-order in gradients of thermal potential and 4-velocity. This happens because such terms stem solely from derivatives of the local equilibrium distribution function, while the irreducible moments $\varrho^{\mu_1\cdots\mu_\ell}_r$ were constructed in terms of the single-particle distribution function, without factorizing its equilibrium component. In order to identify the Navier-Stokes-like terms in the equations, the irreducible moments must be separated into their equilibrium and non-equilibrium parts. In particular, we remark that only the equations of motion for the irreducible moments with $\ell \leq 2$ have non-zero Navier-Stokes-like terms -- these terms vanish for all moments of rank $\ell \geq 3$, as a consequence of Eq.~\eqref{eq:vanishing_integrals}. Nevertheless, we remark that these terms were explicitly calculated in Ref.~\cite{dnmr}. 

In order to obtain an expression for $f_{\mathbf{k}}$, it is necessary to compute the dynamics of the irreducible moments, which in turn satisfy a hierarchy of coupled differential equations. In particular, as previously discussed, the \textit{exact} single-particle distribution function is obtained when the sums in Eq.~\eqref{eq:usual_moments} are taken to infinity, i.e., when the dynamics of \textit{all} its irreducible moments is taken into account. In practice, however, we are required to truncate the expansion in Eq.~\eqref{eq:usual_moments}, including the number of moments required for the series to converge. Nevertheless, the goal of this paper will be different. We will not evaluate the distribution function itself but will instead verify how one can effectively truncate the moment equations derived above and how well this truncation procedure converges. This will be discussed in detail in the next section in the context of boost-invariant expanding systems, where the moment equations derived above can be solved to a very high order.

\section{Gas of massless particles in Bjorken flow}
\label{sec:bjorken}

Bjorken flow \cite{Bjorken1983} is a simplified framework to study ultrarelativistic heavy-ion collisions. It consists in assuming that the bulk matter produced after the collision displays an axial symmetry and is invariant under Lorentz boosts along the collision axis. This highly symmetric flow profile provides the simplest setup in which the fluid-dynamical equations admit analytical solutions \cite{Denicol:2017lxn, Denicol:2019lio, Aniceto:2024pyc} and further serves as a starting point to investigate solutions of the Boltzmann equation \cite{Heinz:2015gka, Jaiswal:2021uvv, Chen:2023vrk, Nugara:2023eku}. 

The spacetime in Bjorken flow is more conveniently described using hyperbolic coordinates, with a geometry defined by the following metric tensor
\begin{equation}
g_{\mu\nu} =  \mathrm{diag} \, (g_{\tau\tau}, g_{xx}, g_{yy}, g_{\eta_s \eta_s}) = \mathrm{diag} \, (1, -1, -1, -\tau^2),
\end{equation}
where $\tau$ is the proper time and $\eta_s$ is the spacetime rapidity. These coordinates are related to the usual Cartesian coordinates through
\begin{equation}
\tau = \sqrt{t^2-z^2}, \hspace{.5cm} \eta_s = \frac{1}{2} \mathrm{ln} \left( \frac{t+z}{t-z} \right).
\end{equation}
In this coordinate system, the only non-zero Christoffel symbols are
\begin{equation}
\Gamma^\tau_{\eta_s \eta_s} = \tau, \hspace{.3cm} \Gamma^{\eta_s}_{\tau \eta_s} = \Gamma^{\eta_s}_{\eta_s \tau} = \frac{1}{\tau}.
\end{equation}
Therefore, all usual derivatives must be replaced by covariant derivatives in the equations of motion for the irreducible moments. In addition, we make the following set of assumptions:
\begin{enumerate}
\item The system is symmetric under reflections with the respect to the $\eta_s$ axis, i.e., $\eta_s \rightarrow - \eta_s$.
\item The system is homogeneous, i.e., invariant under translations, along the $\eta_s$ axis. This implies the system is boost-invariant and thus all fluid-dynamical quantities depend solely on the proper time $\tau$.
\item The system is homogeneous and isotropic (invariant under translations and rotations) in the transverse $xy$-plane.
\end{enumerate}

These assumptions combined lead to a static fluid velocity in hyperbolic coordinates, $u^\mu = (u^\tau, u^x, u^y, u^{\eta_s}) = (1,0,0,0)$. Naturally, the fluid itself is not static, as its expansion is embedded in the metric tensor and thus manifests itself via the covariant derivatives of the 4-velocity \cite{denicolbook}. In particular, the expansion rate and shear tensor are given by
\begin{equation}
\theta = D_\mu u^\mu = \frac{1}{\tau}, \hspace{.5cm} \sigma_{\mu\nu} = D_{\langle\mu}u_{\nu\rangle} = \left( 0, \frac{1}{3\tau}, \frac{1}{3\tau}, - \frac{2\tau}{3} \right).
\end{equation}

Finally, the symmetry assumptions imply that the space-time dependence of $f_{\mathbf{k}}$ is restricted to the time coordinate $\tau$. On top of that, the assumption of isotropy in the $xy$-plane implies that the momentum dependence of $f_{\mathbf{k}}$ can be fully determined by magnitude of the transverse momentum, $k_\bot \equiv \sqrt{k_x^2 + k_y^2}$, and the longitudinal component, $k_{\eta_s}$.  In particular, given the on-shell condition, $k_\mu k^\mu = 0$, $k_\bot$ can be expressed solely in terms of $k_0$ and $k_{\eta_s}$, via, 
$ k_0 = \sqrt{k_\bot^2 + k_{\eta_s}^2/\tau^2 }$. Then, in Bjorken flow, the single-particle distribution function can be written as,
\begin{equation}
f_{\mathbf{k}} = f (\tau, k_0, k_{\eta_s}). \label{f_bjorken0}
\end{equation}
At this point, it is convenient to define a normalized space-like 4-vector $z_\mu = (0, 0, 0, -\tau)$, such that $z_\mu z^\mu = -1$, that is orthogonal to the fluid 4-velocity, $u_\mu z^\mu = 0$, so that Eq.~\eqref{f_bjorken0} can be expressed in a covariant form,
\begin{equation}
f_{\mathbf{k}} = f (\tau, u^\mu k_\mu, z^\mu k_\mu). \label{f_bjorken1}
\end{equation}

The irreducible moments $\varrho^{\mu_1 \cdots \mu_\ell}_r$ are defined as integrals of $f_{\mathbf{k}}$ over momentum space. Thus, in Bjorken flow, they can only depend on $\tau$, $u^\mu$, and $z^\mu$, see Eqs.~\eqref{eq:def_varvarrho} and \eqref{f_bjorken1}. In particular, their tensor structure must be constructed solely from combinations of $u^\mu$, $z^\mu$, and the metric tensor, $g^{\mu\nu}$. The only combination of these tensors that form an irreducible tensor of rank $\ell$ is $z^{\langle \mu_1} \cdots z^{\mu_\ell \rangle}$ \cite{denicolbook}. Thus, the irreducible moments must have the following general form in Bjorken flow, 
\begin{equation}
\varrho^{\mu_1 \cdots \mu_\ell}_n = \mathcal{F}(\tau) z^{\langle \mu_1} \cdots z^{\mu_\ell \rangle},
\end{equation}
where we identify,
\begin{equation}
\mathcal{F}(\tau) = (-1)^\ell \frac{(2\ell-1)!!}{\ell!} z_{\langle \mu_1} \cdots z_{\mu_\ell \rangle} \varrho^{\mu_1 \cdots \mu_\ell}_n,
\end{equation}
making use of the identity \cite{denicolbook}
\begin{equation}
z_{\langle \mu_1} \cdots z_{\mu_\ell \rangle} z^{\langle\mu_1}\cdots z^{\mu_\ell \rangle} = (-1)^\ell \frac{\ell!}{(2\ell-1)!!}.
\end{equation}
The next step it to obtain the explicit form of $\mathcal{F}(\tau)$. First, we note that \cite{denicolbook}
\begin{eqnarray}
z_{\langle \mu_1} \cdots z_{\mu_\ell \rangle} k^{\langle \mu_1} \cdots k^{\mu_\ell \rangle} 
& = & \sum_{q=0}^{\left[\ell/2\right]} C(\ell,q) (-E_{\mathbf{k}}^2)^q \, (-1)^q \, \left( \frac{k_{\eta_s}}{\tau} \right)^{\ell-2q} \notag \\
& = & E_\mathbf{k}^\ell \sum_{q=0}^{\left[\ell/2\right]} C(\ell,q) (\cos\Theta)^{\ell-2q},
\end{eqnarray}
where we have defined $\cos\Theta \equiv k_{\eta_s}/(\tau k_0)$. In particular, we remark that the Legendre polynomials can be expressed as \cite{gradshteyn2014table}
\begin{equation}
P_n (x) = \frac{(2n-1)!!}{n!} \, \sum_{q=0}^{[n/2]} C(n,q) \, x^{n-2q},
\end{equation}
thus leading to
\begin{equation}
z_{\langle \mu_1} \cdots z_{\mu_\ell \rangle} k^{\langle \mu_1} \cdots k^{\mu_\ell \rangle} = \frac{\ell!}{(2\ell-1)!!} \, E_{\mathbf{k}}^\ell \, P_\ell(\cos\Theta).
\end{equation}
Wherefore, it follows that
\begin{eqnarray}
z_{\langle \mu_1} \cdots z_{\mu_\ell \rangle} \varrho_n^{\mu_1 \cdots \mu_\ell} = \frac{\ell!}{(2\ell-1)!!} \int dK k_0^{n+\ell} P_\ell(\cos\Theta) f_{\mathbf{k}}.
\end{eqnarray}

The system's invariance under reflections around the $\eta_s$-axis further implies that $f (\tau, k_0, k_{\eta_s}) = f (\tau, k_0, - k_{\eta_s})$. Moreover, the Legendre polynomials, $P_\ell(\cos\Theta)$, are even (odd) functions of $k_{\eta_s}$, as long as $\ell$ is equally even (odd). Therefore, irreducible moments of odd rank are identically zero in Bjorken flow \cite{denicolbook}. Taking $\ell \rightarrow 2\ell$, we have
\begin{equation}
z_{\langle \mu_1} \cdots z_{\mu_{2\ell} \rangle} \varrho_n^{\mu_1 \cdots \mu_{2\ell}} \equiv \frac{(2\ell)!}{(4\ell-1)!!} \varrho_{n, \ell}, 
\end{equation}
where we have defined the new fields\footnote{We remark that the moments $\mathcal{L}_\ell$, investigated in Refs.~\cite{Blaizot:2017lht, Blaizot:2017ucy, Blaizot:2021cdv, Brewer:2022ifw, Jaiswal:2022udf}, are a subset of the irreducible moments $\varrho_{n, \ell}$ for $n=2$. Furthermore, what these references refer to as mode coupling theory is equivalent to the traditional method of moments discussed here. }
\begin{equation}
\label{eq:rho-bjorken}
\varrho_{n, \ell} = \int dK k_0^n P_{2\ell}(\cos\Theta) f_{\mathbf{k}}.
\end{equation}
In particular, taking $f_{\mathbf{k}} = f_{0\mathbf{k}}$, we obtain the equilibrium value of the irreducible moments, which vanish unless $\ell = 0$, as a consequence of the orthogonality of the Legendre polynomials \cite{gradshteyn2014table},
\begin{equation}
\varrho_{n, \ell}^{\mathrm{eq}} = g \, e^\alpha \frac{(n+1)!}{2 \pi^2}T^{n+2} \delta_{\ell 0},
\end{equation}
with $g$ being the degeneracy factor, which we have previously assumed to be $1$. In summary, the irreducible moments of a generic single-particle distribution function in Bjorken flow can be written as
\begin{equation}
\varrho^{\mu_1 \cdots \mu_{2\ell}}_n = \varrho_{n + 2\ell, \ell} z^{\langle \mu_1} \cdots z^{\mu_{2\ell} \rangle}. \label{eq:mom_bjorken}
\end{equation}

In order to obtain the equation of motion satisfied by $\varrho_{n,\ell}$, it is still necessary to provide an expression for the collision term, $C[f]$. In general, obtaining a closed expression for $C[f]$ is the most challenging part of solving the Boltzmann equation with the method of moments. It typically requires the computation of several integrals involving the single-particle distribution function and the transition rate that determines the scattering processes, which can be rather cumbersome even in the linear regime \cite{Denicol:2022bsq, Wagner:2023joq}. In this work, however, we adopt a rather simple prescription for the collision term, known as the relaxation time approximation \cite{oldRTA}. In this approach, the single-particle distribution function is assumed to relax to its equilibrium value within a timescale $\tau_R$, and the collision term becomes simply
\begin{equation}
\label{eq:RTA}
C[f] = - \frac{E_{\mathbf{k}}}{\tau_R} \left( f_{\mathbf{k}} - f_{0 \mathbf{k}} \right) \Longrightarrow \mathcal{C}^{\langle \mu_1 \cdots \mu_\ell \rangle}_{r - 1} = - \frac{1}{\tau_R} \left( \rho^{\mu_1 \cdots \mu_\ell}_r - \rho^{\mu_1 \cdots \mu_\ell}_{r, \, \mathrm{eq}} \right).
\end{equation}
Hence, all irreducible moments also evolve towards their equilibrium values within the same timescale. In particular, we assume this to be the shear relaxation time, $\tau_R = \tau_\pi = 5\eta/(\varepsilon_0+P_0)$ \cite{dnmr}. In order to be consistent with the conservation of energy and momentum, this prescription requires the imposition of Landau matching conditions \cite{nRTA}, which defines the values of the temperature and chemical potential out of equilibrium so that the particle and energy density are fixed to their equilibrium values, $\varrho_{1,0} \equiv \varrho_{1,0}^{\mathrm{eq}}$ and $\varrho_{2,0} \equiv \varrho_{2,0}^{\mathrm{eq}}$, respectively.

We are now finally in position to obtain a set of coupled equations of motion for the irreducible moments in the framework of Bjorken flow. Replacing Eqs.~\eqref{eq:mom_bjorken} and Eq.~\eqref{eq:RTA} into Eq.~\eqref{eq:eom_moms} and contracting it with $z_{\langle \mu_1} \cdots z_{\mu_{2\ell} \rangle}$, we obtain
\begin{eqnarray}
\label{eq:eom-rhos-bjorken}
D_\tau \varrho_{n, \ell} & = & - \frac{1}{\tau_R} \left( \varrho_{n, \ell} - \varrho_{n, \ell}^{\mathrm{eq}} \right) - \mathcal{P} \frac{\varrho_{n, \ell - 1}}{\tau} - \mathcal{Q} \frac{\varrho_{n, \ell}}{\tau} - \mathcal{R} \frac{\varrho_{n, \ell + 1}}{\tau},
\end{eqnarray}
where we have introduced the following coefficients
\begin{subequations}
\begin{align}
\mathcal{P} &= 2 \ell \frac{(n + 2\ell ) (2\ell - 1)}{(4\ell + 1) (4\ell -1 )}, \\
\mathcal{Q} &=  \frac{2 \ell (2\ell + 1) + n (24\ell^2 + 12 \ell - 3)}{3 (4\ell - 1) (4\ell + 3)} + \frac{2}{3}, \\
\mathcal{R} &= (n - 2 \ell -1) \frac{(2\ell + 1) (2\ell + 2)}{(4\ell + 1) (4\ell + 3)}.
\end{align}
\end{subequations}
We recovered the set of differential equations previously obtained in Ref.~\cite{denicolbook}, where the method of moments was directly constructed assuming the symmetries of Bjorken flow within the relaxation time approximation. In the present work, however, we followed a different path: we constructed the method of moments for a general flow configuration and only then employed the assumptions of Bjorken flow and the relaxation time approximation. Finally, contracting the equation of motion for $\varrho^{\mu_1 \cdots \mu_{2\ell}}_n$ with $z_{\langle \mu_1} \cdots z_{\mu_\ell \rangle}$, we obtained the same result as in Ref.~\cite{denicolbook}, showing the consistency of the calculations developed here. 

We note that the DNMR equations of fluid dynamics \cite{dnmr} appear when the moment equations \eqref{eq:eom_moms} are truncated at rank $2$, i.e., all moments of rank higher than $2$ are neglected. Moreover, the inclusion of irreducible moments of rank 3 and 4 in the hierarchy of equations (in Bjorken flow, this consists in including irreducible moments $\varrho_{n, 2}$) lead to third-order fluid dynamics, as shown in Ref.~\cite{deBrito:2023tgb}. As a matter of fact, if a sufficiently large number of moments is included in the hierarchy, one expects to obtain the exact solution for the hydrodynamic fields. However, the behavior and convergence of these solutions have not been fully explored. The next step is to thoroughly analyze the behavior of the irreducible moments and the convergence of this hierarchy of differential equations. 

\section{Convergence of the solutions}
\label{sec:convergence}

We solve Eqs.~\eqref{eq:eom-rhos-bjorken} by imposing a truncation of the moment expansion. In practice, this means that irreducible moments of a given rank (or higher) will simply be neglected in the calculations. In the highly symmetric flow configuration considered in the previous section, this is implemented by taking $\varrho_{n,\ell} = 0$, $\forall \,\, \ell > \ell_{\mathrm{max}}$, with the parameter $\ell_{\mathrm{max}}$ quantitatively specifying our truncation. Once this is done, Eqs.~\eqref{eq:eom-rhos-bjorken} can be solved using the Runge-Kutta algorithm. Unless stated otherwise, we consider the system to be in equilibrium at an initial time $\tau_0$, with a temperature $T(\tau_0) = 1$ GeV and a vanishing chemical potential. Also, we consider three distinct values of initial time, always specified in units of $\tau_R$, $\tau_0/\tau_R = 0.01$, $0.1$ and $1$. 

\begin{figure}[ht]
\centering
\begin{subfigure}{0.31\textwidth}
  \includegraphics[width=\linewidth]{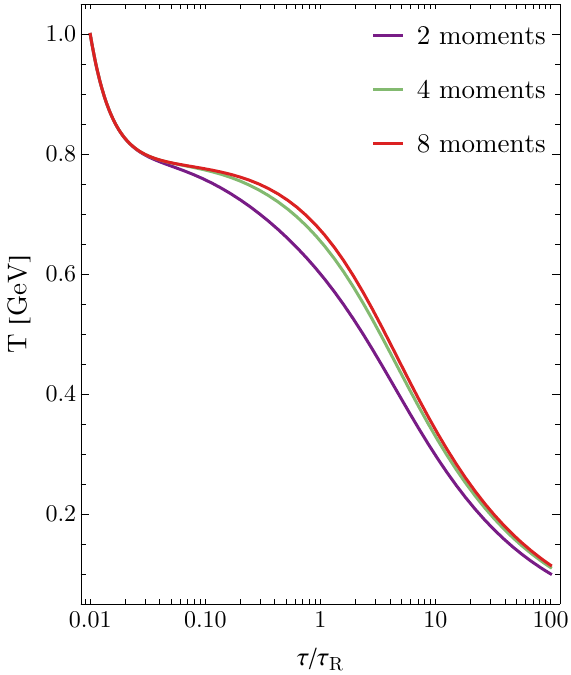}
\end{subfigure}\hfil
\begin{subfigure}{0.31\textwidth}
  \includegraphics[width=\linewidth]{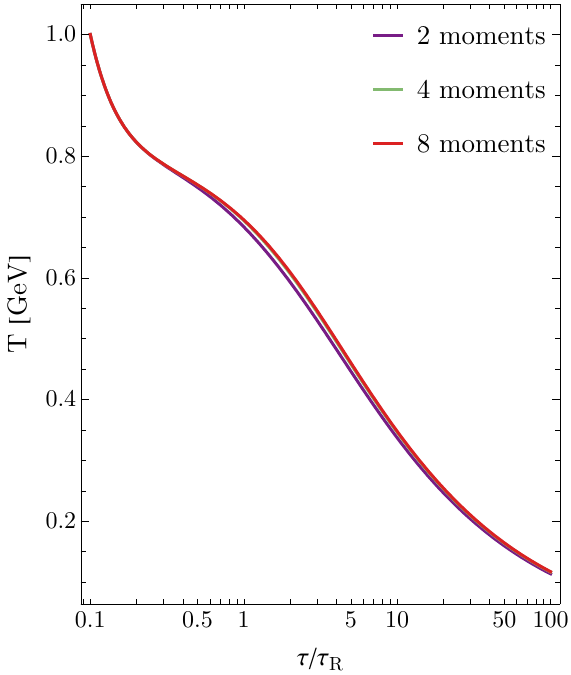}
\end{subfigure}\hfil
\begin{subfigure}{0.31\textwidth}
  \includegraphics[width=\linewidth]{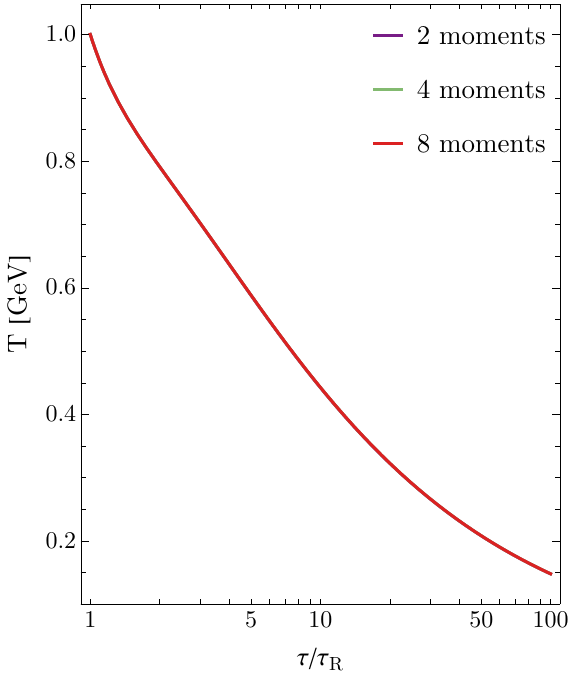}
\end{subfigure}\hfil \\
\begin{subfigure}{0.30\textwidth}
  \includegraphics[width=\linewidth]{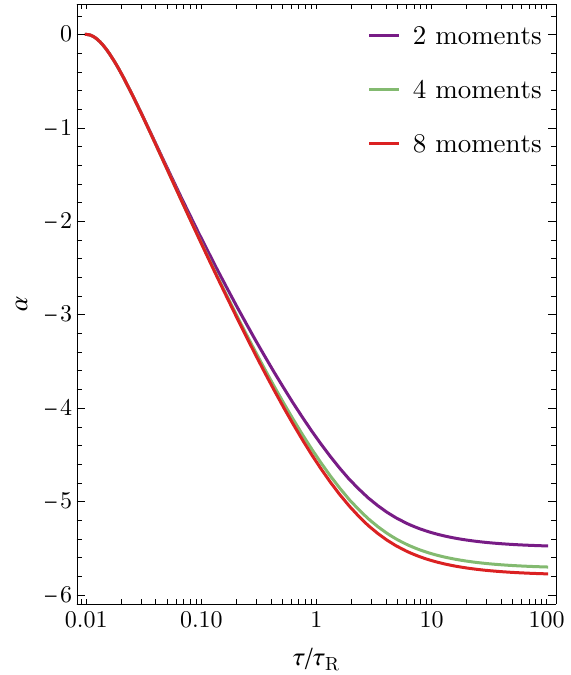}
\end{subfigure}\hfil
\begin{subfigure}{0.30\textwidth}
  \includegraphics[width=\linewidth]{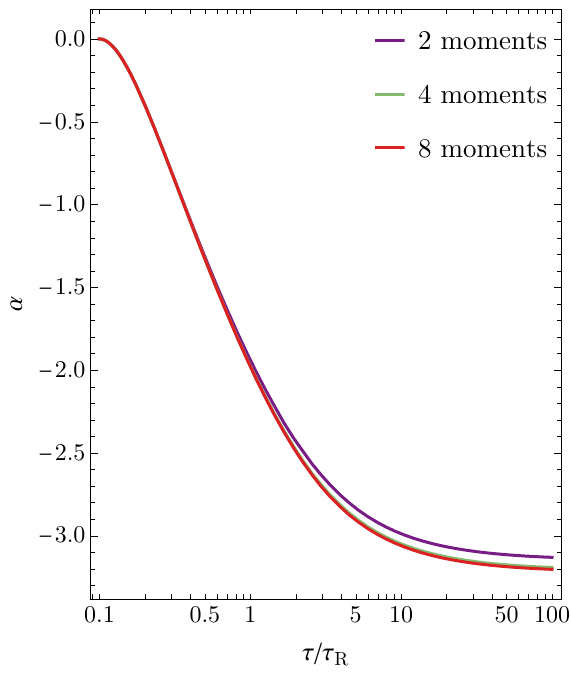}
\end{subfigure}\hfil
\begin{subfigure}{0.30\textwidth}
  \includegraphics[width=\linewidth]{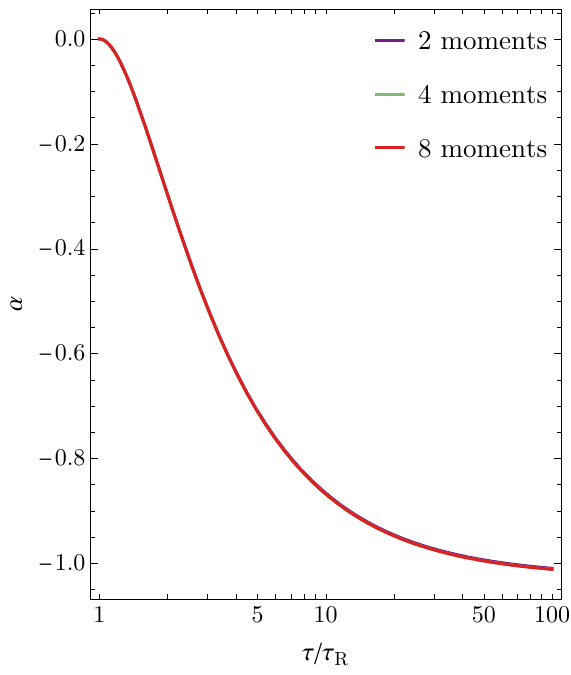} 
\end{subfigure}\hfil
\caption{(Color online) Solutions for the temperature (upper panels) and thermal potential (lower panels) considering different truncations and values for the initial rescaled time and assuming $T(\tau_0) = 1$ GeV and $n(\tau_0) = T(\tau_0)^3/\pi^2$.}
\label{fig:therm-var}
\end{figure}

We first look at the convergence of solutions for the temperature and thermal potential as a function of the rescaled time, $\tau/\tau_R$, for $\ell_{\mathrm{max}} = 2$, $4$ and 8. These quantities are obtained by solving Eqs.~\eqref{eq:eom-rhos-bjorken} and then using Landau matching conditions to extract $T$ and $\alpha$ from the particle number and energy densities. In Fig.~\ref{fig:therm-var}, we display solutions for the temperature (upper panels) and thermal potential (lower panels) as functions of the rescaled time. We note that all the dependence on $\eta/s$ is embedded in the relaxation time, $\tau_R$, and, by  plotting our results as a function of the rescaled time,
the magnitude of the dissipative  effects is solely determined by the initial value of $\tau/\tau_R$ -- that is, more dissipative systems can be probed by considering smaller values for the initial rescaled time.
We observe that, as the initial rescaled time is smaller, i.e. as the system becomes more dissipative, more moments have to be included in the hierarchy of differential equations \eqref{eq:eom-rhos-bjorken} in order for the solutions to converge. 

\begin{figure}[ht]
\centering
\begin{subfigure}{0.31\textwidth}
  \includegraphics[width=\linewidth]{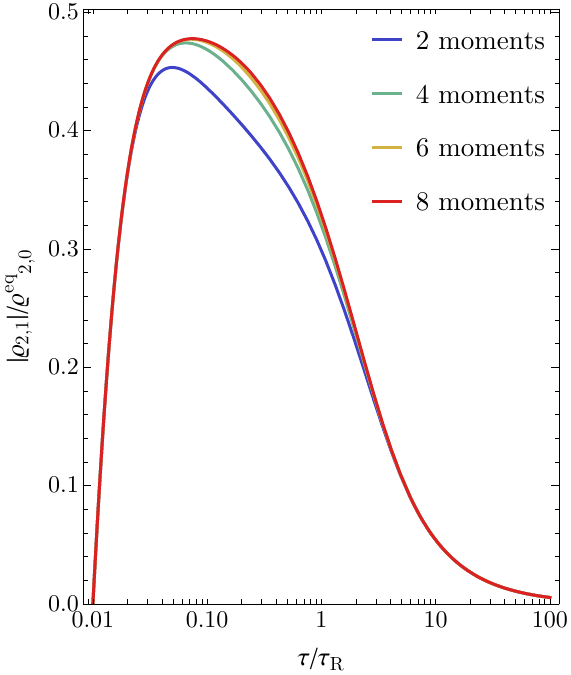}
\end{subfigure}\hfil
\begin{subfigure}{0.31\textwidth}
  \includegraphics[width=\linewidth]{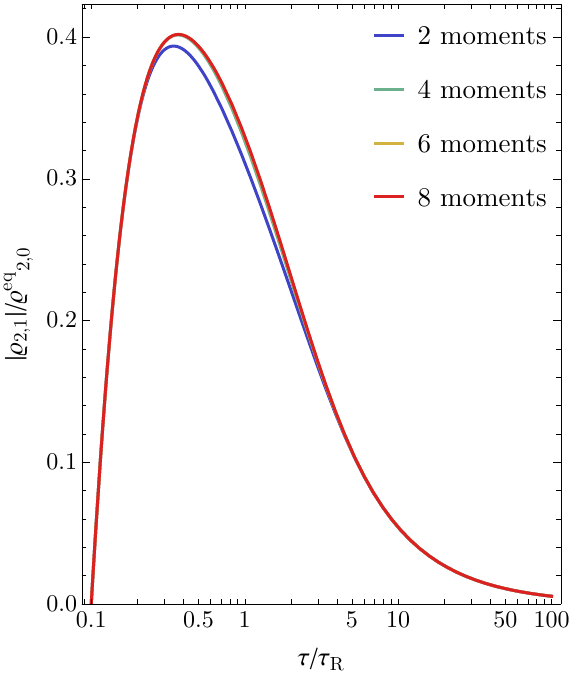}
\end{subfigure}\hfil
\begin{subfigure}{0.31\textwidth}
  \includegraphics[width=\linewidth]{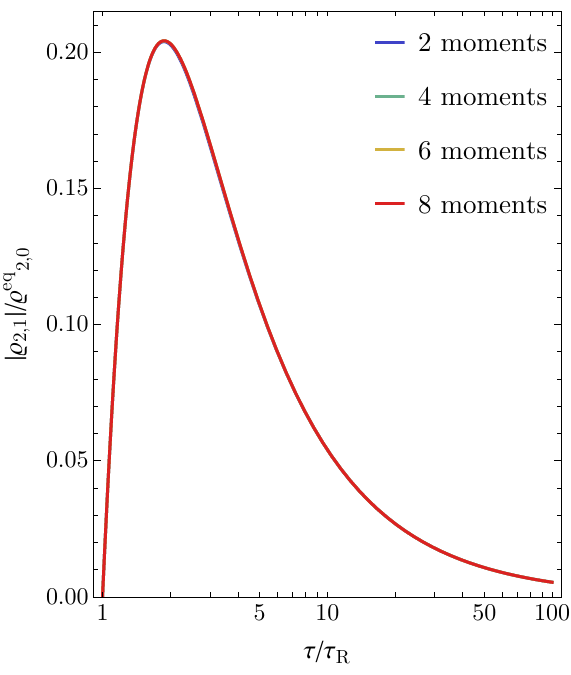}
\end{subfigure}\hfil \\
\begin{subfigure}{0.30\textwidth}
  \includegraphics[width=\linewidth]{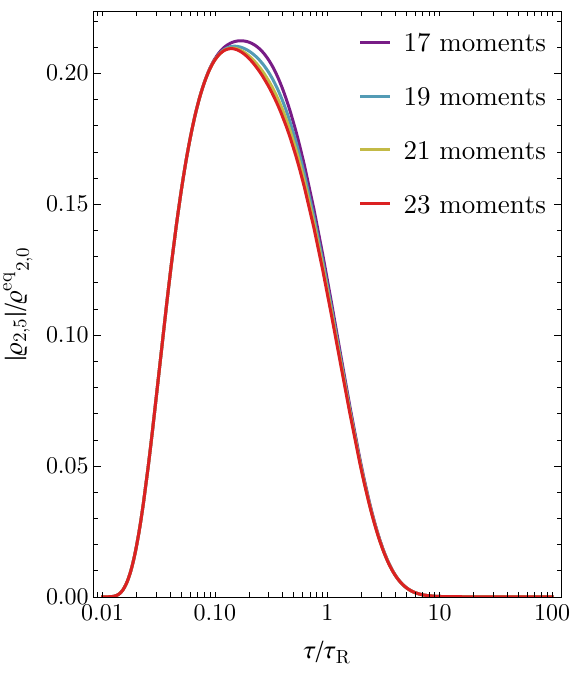}
\end{subfigure}\hfil
\begin{subfigure}{0.30\textwidth}
  \includegraphics[width=\linewidth]{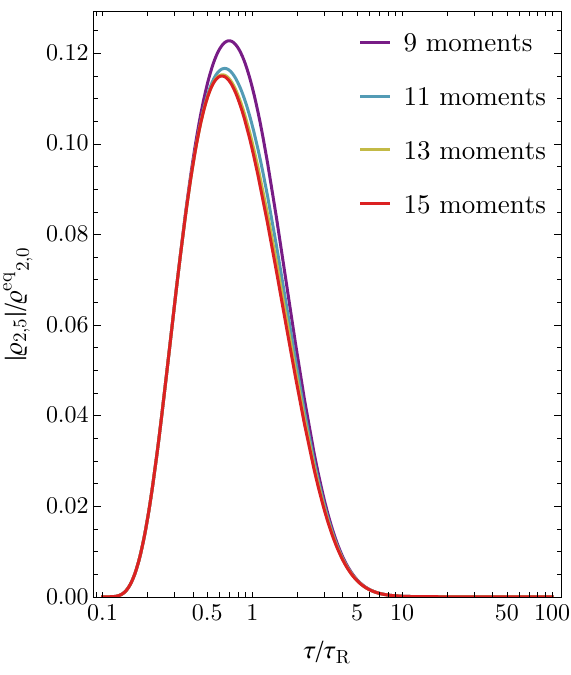}
\end{subfigure}\hfil
\begin{subfigure}{0.30\textwidth}
  \includegraphics[width=\linewidth]{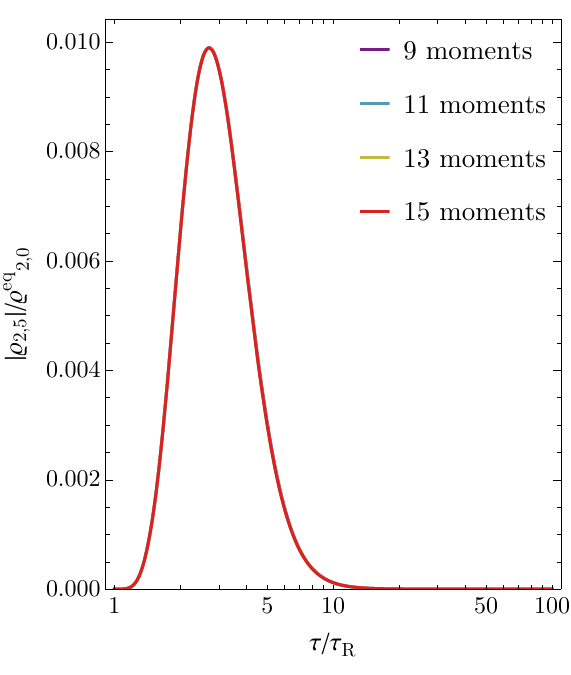} 
\end{subfigure}\hfil \\
\begin{subfigure}{0.30\textwidth}
  \includegraphics[width=\linewidth]{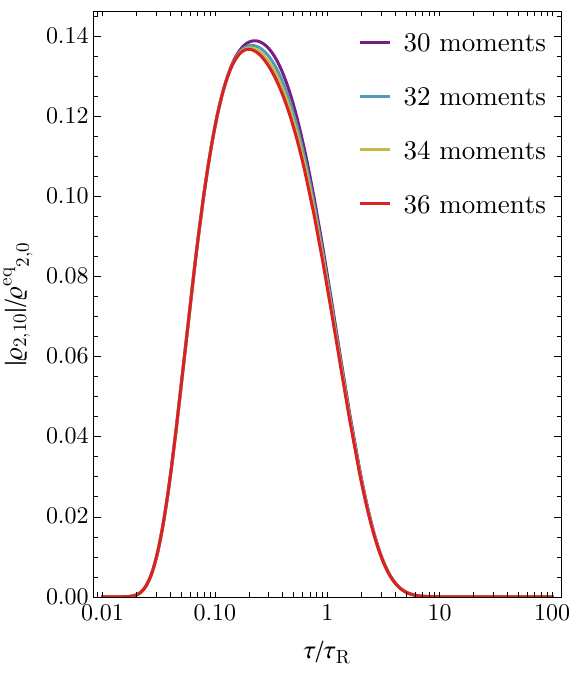}
\end{subfigure}\hfil
\begin{subfigure}{0.30\textwidth}
  \includegraphics[width=\linewidth]{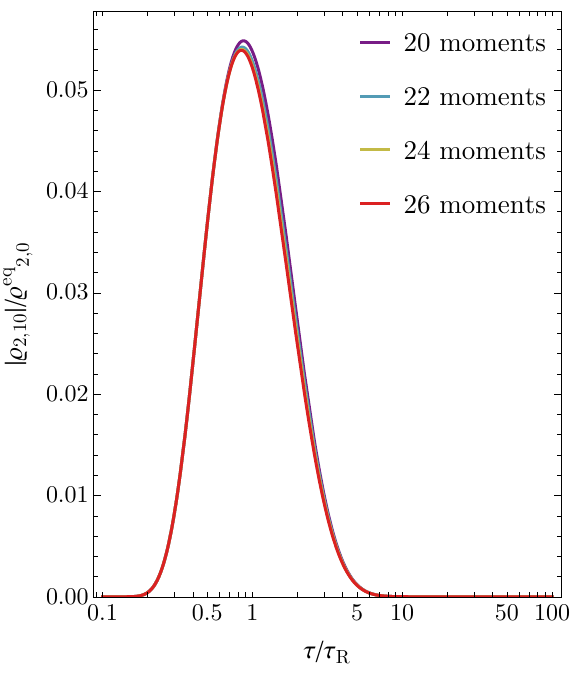}
\end{subfigure}\hfil
\begin{subfigure}{0.30\textwidth}
  \includegraphics[width=\linewidth]{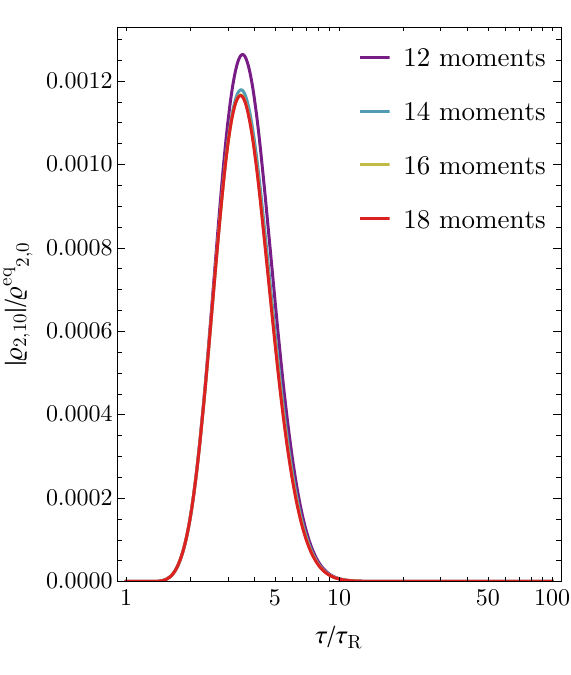} 
\end{subfigure}\hfil
\caption{(Color online) Solutions for $\varrho_{2,1}/\varrho_{2,0}^\mathrm{eq}$ (upper panels), $\varrho_{2,5}/\varrho_{2,0}^\mathrm{eq}$ (middle panels) and $\varrho_{2,10}/\varrho_{2,0}^\mathrm{eq}$ (lower panels) considering different truncations and values for the initial rescaled time and assuming $T(\tau_0) = 1$ GeV and $n(\tau_0) = T(\tau_0)^3/\pi^2$.}
\label{fig:moments-dif-trunc}
\end{figure}

Next we look at the convergence of certain irreducible moments. In Fig.~\ref{fig:moments-dif-trunc}, we display $\varrho_{2,1}/\varrho_{2,0}^\mathrm{eq}$ (which is related to the shear-stress tensor as $\pi^{\eta_s \eta_s} = - 2 \varrho_{2,1}/3$ \cite{denicolbook}), $\varrho_{2,5}/\varrho_{2,0}^\mathrm{eq}$ and $\varrho_{2,10}/\varrho_{2,0}^\mathrm{eq}$ as functions of the rescaled time $\tau/\tau_R$, for several values of $\ell_{\mathrm{max}}$. It can readily be seen that, as more moments are included in the hierarchy of differential equations, the solutions for the irreducible moments gradually converge to a unique curve. Moreover, as the initial rescaled time becomes smaller, the irreducible moments become larger in magnitude and more moments have to be included in order to observe convergence, i.e., a larger value of $\ell_{\mathrm{max}}$ is required in order for the solutions to converge. We remark that this behavior is also observed for different values of $n$, which defines the power of energy in Eq.~\eqref{eq:rho-bjorken}, but we do not display these cases for the sake of simplicity.

\begin{figure}[ht]
\centering
\begin{subfigure}{0.30\textwidth}
  \includegraphics[width=\linewidth]{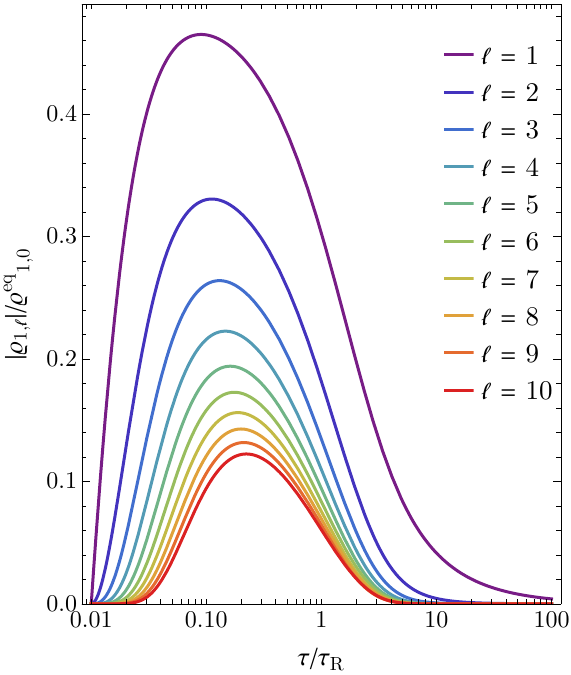}
\end{subfigure}\hfil
\begin{subfigure}{0.30\textwidth}
  \includegraphics[width=\linewidth]{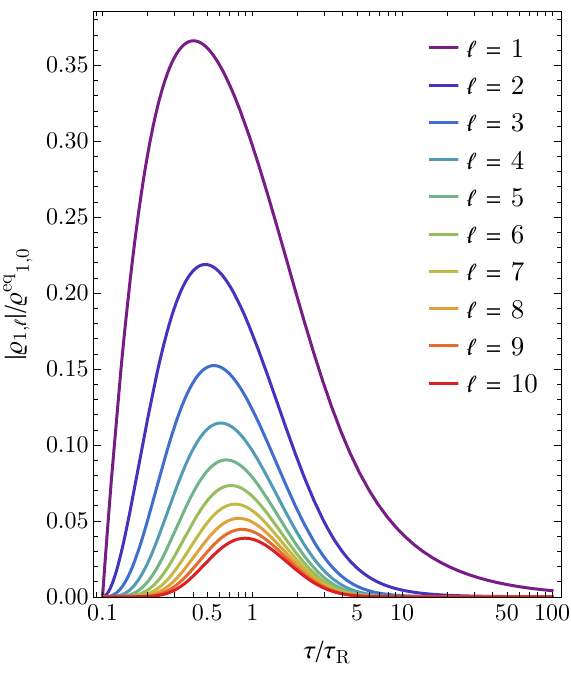}
\end{subfigure}\hfil
\begin{subfigure}{0.30\textwidth}
  \includegraphics[width=\linewidth]{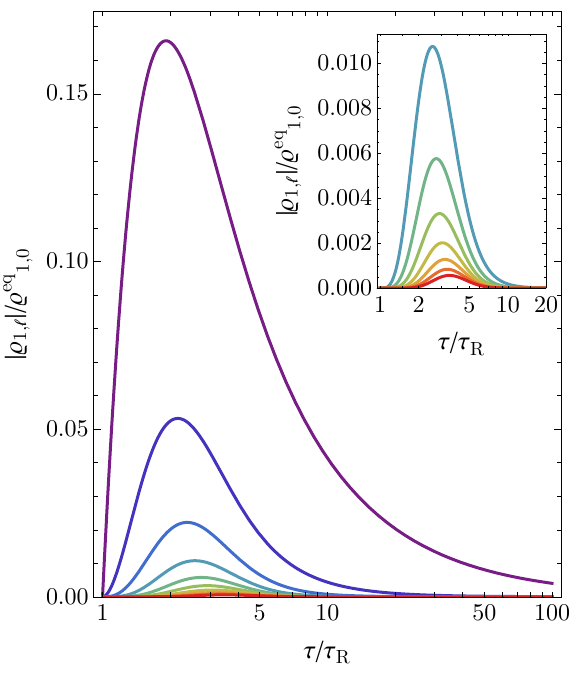} 
\end{subfigure}\hfil 
\caption{(Color online) Normalized irreducible moments for $n=1$ and considering different values for $\ell$ and initial rescaled time and assuming $T(\tau_0) = 1$ GeV and $n(\tau_0) = T(\tau_0)^3/\pi^2$.}
\label{fig:moments-nFixed}
\end{figure}

We now look at the irreducible moments, $\varrho_{n\ell}$, fixing the value of $n$ and varying the parameter $\ell$. We solve Eqs.~\eqref{eq:eom-rhos-bjorken} for $\ell_{\mathrm{max}}=100$ and portray, in Fig.~\ref{fig:moments-nFixed}, the absolute value of the irreducible moments normalized by their equilibrium value, $\vert \rho_{n,\ell} \vert/\rho^{\mathrm{eq}}_{n,0}$, for $n=1$ and $\ell=1$--$10$ as a function of $\tau/\tau_R$. Each panel of Fig.~\ref{fig:moments-nFixed} displays solutions obtained for a different choice of initial time.
We observe that, for a fixed $n$, the normalized irreducible moments become smaller (in magnitude) as the value of $\ell$ is increased. This is consistent with the apparent convergence observed for the solutions of the moment equations obtained with our truncation scheme -- moments become smaller as the value of $\ell$ is increased and, for a sufficiently large value of $\ell$, it becomes a good approximation to simply set them to zero. Naturally, the moments with $\ell \ll \ell_{\mathrm{max}}$ are well approximated in this scheme, whereas those with $\ell \sim \ell_{\mathrm{max}}$ are usually not. This is in agreement with the behavior observed in Figs.~\ref{fig:therm-var} and \ref{fig:moments-dif-trunc}. 

\begin{figure}[ht]
\centering
\begin{subfigure}{0.30\textwidth}
  \includegraphics[width=\linewidth]{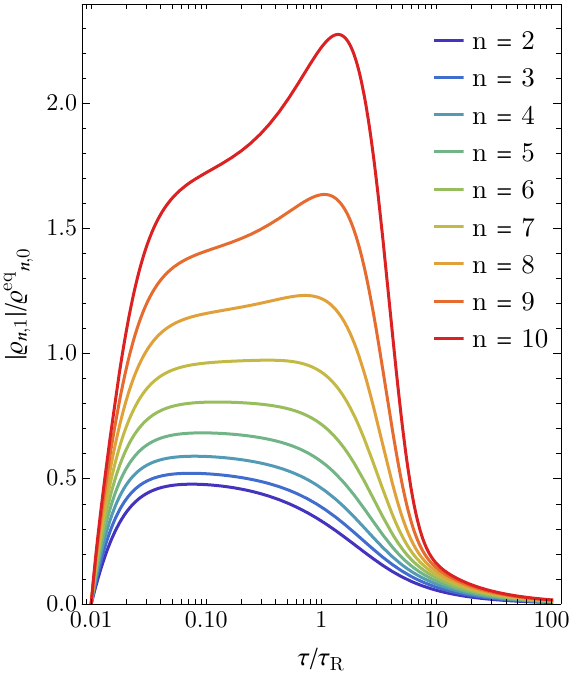}
\end{subfigure}\hfil
\begin{subfigure}{0.30\textwidth}
  \includegraphics[width=\linewidth]{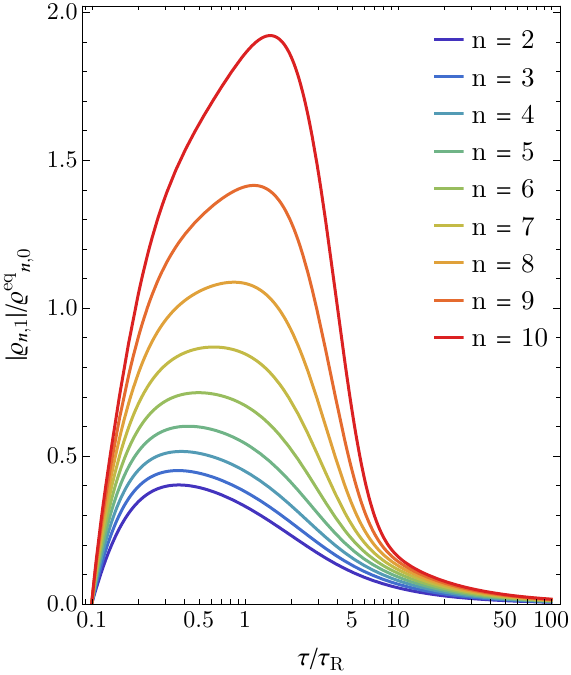}
\end{subfigure}\hfil
\begin{subfigure}{0.30\textwidth}
  \includegraphics[width=\linewidth]{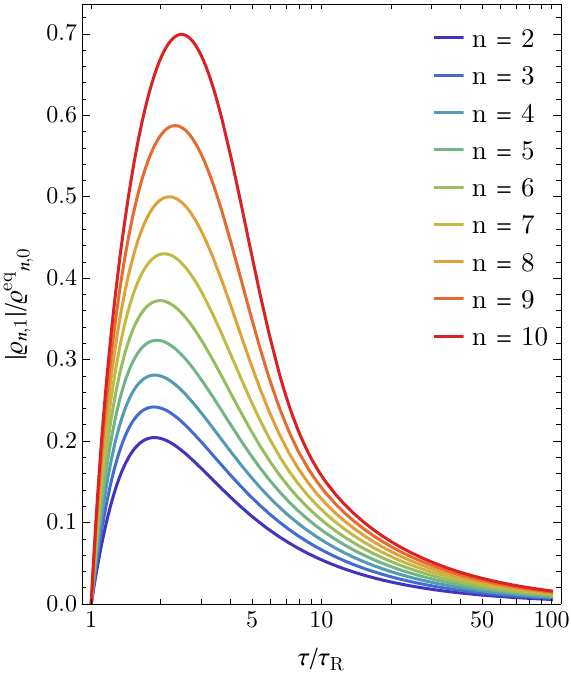} 
\end{subfigure}\hfil 
\caption{(Color online) Normalized irreducible moments for $\ell_{\mathrm{max}}=10$ considering different values for the initial rescaled time and assuming $T(\tau_0) = 1$ GeV and $n(\tau_0) = T(\tau_0)^3/\pi^2$.}
\label{fig:moments}
\end{figure}

Next, we analyze the behavior of the irreducible moments, $\varrho_{n\ell}$, for $\ell=1$ and several values of $n$. We solve Eqs.~\eqref{eq:eom-rhos-bjorken} with $\ell_{\mathrm{max}} = 10$ (which is sufficiently large to ensure the convergence of the solutions) and considering $n=1$--$10$. In Fig.~\ref{fig:moments}, we display the absolute value of the normalized irreducible moments $\vert \varrho_{n1}\vert/\varrho_{n0}^\mathrm{eq}$ as a function of the rescaled time $\tau/\tau_R$. We observe that as the value of $n$ is increased, the magnitude of the normalized moments is also increased, and this behavior becomes even more manifested as the initial rescaled time is smaller. We remark that this behavior is also observed for different choices of $\ell$, but we not display these results here for the sake of simplicity. This behavior of the irreducible moments as one increases the parameter $n$ is significant, since it renders the task of determining the single-particle distribution function complicated. In the method of moments, the single-particle distribution function is expressed in terms of a sum of irreducible moments, cf.~Eq.~\eqref{eq:usual_moments}, and it is generally assumed that this series converges rapidly, at least for relatively small values of momentum. However, here we see that $\varrho_{n\ell}$ grows significantly with $n$ at intermediate (rescaled) times, making it challenging to calculate the moment expansion to a sufficiently high order and establishing its convergence. This will be thoroughly investigated in an upcoming follow-up work.

\begin{figure}[ht]
\centering
\begin{subfigure}{0.4\textwidth}
  \includegraphics[width=\linewidth]{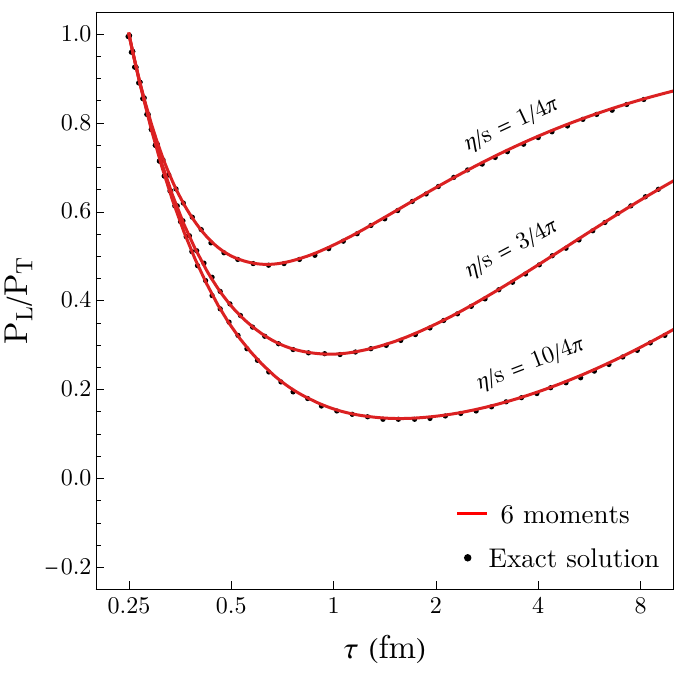}
\end{subfigure}\hfil
\begin{subfigure}{0.4\textwidth}
  \includegraphics[width=\linewidth]{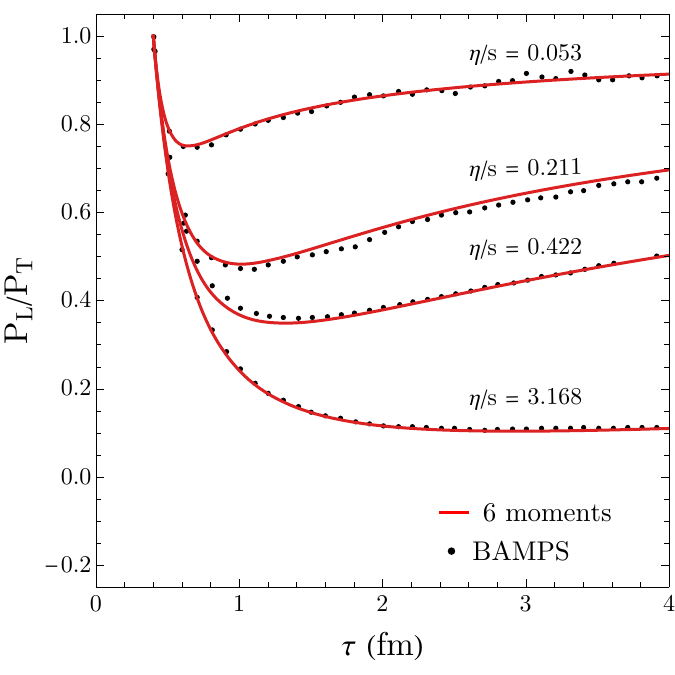}
\end{subfigure}\hfil
\caption{(Color online) Pressure anisotropy for $T(\tau_0) = 300$ MeV and $\tau_0 = 0.25$ fm (left panel) and $T(\tau_0) = 500$ MeV and $\tau_0 = 0.4$ fm (right panel).}
\label{fig:press-aniso}
\end{figure}

Last, we compute the pressure anisotropy considering different values for the shear viscosity over entropy density, $\eta/s$, and initial temperatures calibrated to emulate the hot and dense matter created at the Relativistic Heavy-Ion Collider (RHIC) and the Large Hadron Collider (LHC). In the left panel of Fig.~\ref{fig:press-aniso}, we display the pressure anisotropy as a function of the proper time, calculated from Eqs.~\eqref{eq:eom-rhos-bjorken} with $\ell_{\mathrm{max}} = 6$, and compare them to semi-analytical solutions for this quantity derived in Ref.~\cite{Florkowski:2017ixx}. Three values of shear viscosity over entropy density are considered: $\eta/s=1/(4\pi)$, $3/(4\pi)$, and $10/(4\pi)$ -- here, since we do not display the solution as a function of the rescaled time, the curves will depend on the magnitude of the shear viscosity coefficient. For the sake of comparison, we consider the initial conditions proposed in Ref.~\cite{Florkowski:2017ixx}, with the system being initially at equilibrium at an initial time $\tau_0 = 0.25$ fm, with an initial temperature of $T (\tau_0) = 300$ MeV. For this type of initial condition, we observe that including $6$ moments in the moment equations is sufficient to obtain a convergent solution. In the right panel of Fig.~\ref{fig:press-aniso}, we compare our solutions to numerical solutions of the Boltzmann equation obtained using the BAMPS simulation code \cite{Xu:2004mz, Xu:2007aa}, in Ref.~\cite{El:2009vj}. For the sake of comparison, we now use the initial conditions employed in Ref.~\cite{El:2009vj}, i.e., a system in equilibrium at $\tau=0.4$ fm with an initial temperature of $T=500$ MeV. We note that the solutions calculated with the method of moments are in surprisingly good agreement with the numerical solutions from BAMPS for practically all values of shear viscosity employed. This did not have to be the case, since in our solutions the collision term was simplified significantly by imposing the relaxation time approximation. This indicates that the relaxation time approximation, even though extremely rudimentary, can still provide a reasonable approximation for the pressure anisotropy. Nevertheless, this may be a feature of the highly symmetric flow configuration considered here.

\section{Conclusions and discussion}
\label{sec:conc}

We have derived the equations of motion for all the \textit{irreducible} moments of the single-particle distribution function. As expected from previous less general results \cite{dnmr,deBrito:2023tgb}, the equations of motion for the moments are coupled, with the particular challenging feature of the dynamics of lower-rank moments being always coupled to those of a higher-rank, leading to an endless tower of equations. In this work we then investigated how this hierarchy of equations can be properly truncated and solved. This is a fundamental step in making it possible to solve the Boltzmann equation using the method of moments. 

We have specifically investigated the solutions for the irreducible moments considering a gas of classical massless particles undergoing a Bjorken expansion, assuming the relaxation time approximation for the collision term \cite{oldRTA}. In this flow profile, the dynamics of the irreducible moments is entirely contained in the scalars $\varrho_{n \ell}$, which then satisfy a simplified set of coupled differential equations of motion. We observe that, as we increase the truncation of the hierarchy of differential equations for the irreducible moments, Eqs.~\eqref{eq:eom-rhos-bjorken}, i.e., as the dynamics of more moments is taken into account, their solutions gradually converge. In particular, we observe that for more dissipative systems, the irreducible moments increase in absolute value, as expected, and even more moments must be included in order for the solutions to converge. In addition, we computed the pressure anisotropy taking $\ell_{\mathrm{max}}=6$ and obtained results in rather good agreement with numerical solutions of the Boltzmann equation for initial conditions calibrated to reproduce the hot and dense matter created at RHIC and LHC considering a wide range of values for the shear viscosity over entropy density. This serves as an important test for the capability of the method of moments in providing accurate solutions for moments of the single-particle distribution function.

Finally, we observed that the magnitude of the moments $\rho_{n\ell}$ always increases with increasing $n$. This makes it a challenge to calculate the single-particle distribution function using the method of moments, since higher-order terms in the moment expansion will not be necessarily small. Understanding the convergence of this series is an important task and will be the primary scope of an upcoming follow-up work.

\begin{acknowledgments}
C. V. P. B. is partly funded by Coordenação de Aperfeiçoamento de Pessoal de Nível Superior (CAPES), Finance Code 001, Award No. 88881.722616/2022-01 and by Conselho Nacional de Desenvolvimento Científico e Tecnológico (CNPq), Grant No. 140453/2021-0. G. S. D. also acknowledges CNPq as well as Fundação Carlos Chagas Filho de Amparo à Pesquisa do Estado do Rio de Janeiro (FAPERJ), Grant No.~E-26/202.747/2018. 
\end{acknowledgments}

\appendix


\section{Derivation of the equation of motion for the irreducible moments}
\label{app_calculations}

In this appendix, we outline the details of the derivation of the equation of motion for the irreducible moments of arbitrary rank, given in Eq.~\eqref{eq:eom_moms}. The starting point is
\begin{equation}
\dot{\varrho}^{\langle\mu_1\cdots\mu_\ell\rangle}_r=\Delta^{\mu_1\cdots\mu_\ell}_{\nu_1\cdots\nu_\ell}\left( \left\langle \frac{dE^r_{\mathbf{k}}}{d\tau} k^{\langle\nu_1}\cdots k^{\nu_\ell\rangle} \right\rangle + \left\langle  E^r_{\mathbf{k}} \frac{d}{d\tau}\left(k^{\langle\nu_1}\cdots k^{\nu_\ell\rangle}\right) \right\rangle + \left\langle E^r_{\mathbf{k}} k^{\langle\nu_1}\cdots k^{\nu_\ell\rangle} \frac{d f_{\mathbf{k}}}{d\tau} \right\rangle \right). \label{eq2}
\end{equation}
where we have employed the notation introduced in Ref.~\cite{dnmr}, $\langle \cdots\rangle \equiv \int dK (\cdots)f_{\mathbf{k}}$. In the following, we compute each term in the right-hand side individually.

\subsection{First term}

The first term is
\begin{equation}
\Delta^{\mu_1\cdots\mu_\ell}_{\nu_1\cdots\nu_\ell} \left\langle \frac{dE^r_{\mathbf{k}}}{d\tau} k^{\langle\nu_1}\cdots k^{\nu_\ell\rangle} \right\rangle,
\end{equation}
where
\begin{equation}
\frac{dE^r_{\mathbf{k}}}{d\tau} = r E^{r-1}_{\mathbf{k}} k^{\langle\mu\rangle} \dot{u}_\mu,
\end{equation}
since the 4-momentum does not depend on spacetime. Furthermore, it is possible to replace $k^\mu$ by its orthogonal projection with respect to the 4-velocity, $k^{\langle \mu \rangle}$, since the 4-acceleration of the fluid is orthogonal to its 4-velocity, $u_\mu \dot{u}^\mu=0$. Therefore,
\begin{equation}
\Delta^{\mu_1\cdots\mu_\ell}_{\nu_1\cdots\nu_\ell} \left\langle \frac{dE^r_{\mathbf{k}}}{d\tau} k^{\langle\nu_1}\cdots k^{\nu_\ell\rangle} \right\rangle = r \dot{u}_{\nu_{\ell+1}} \Delta^{\mu_1\cdots\mu_\ell}_{\nu_1\cdots\nu_\ell} \left\langle  E^{r-1}_{\mathbf{k}} k^{\langle\nu_1}\cdots k^{\nu_\ell\rangle}k^{\langle\nu_{\ell+1}\rangle} \right\rangle. \label{eq11}
\end{equation}
In order to obtain a result in terms of irreducible moments, it is necessary to compute the irreducible decomposition of the term $k^{\langle\nu_1}\cdots k^{\nu_\ell\rangle}k^{\langle\nu_{\ell+1}\rangle}$. First, we write
\begin{equation}
k^{\langle\nu_1}\cdots k^{\nu_\ell\rangle}k^{\langle\nu_{\ell+1}\rangle} = \Delta_{\alpha_1 \cdots \alpha_\ell}^{\nu_1 \cdots \nu_\ell}\Delta^{\nu_{\ell+1}}_{\alpha_{\ell+1}}k^{\langle\alpha_{1}\rangle}\cdots k^{\langle\alpha_{\ell+1}\rangle}.
\end{equation}
Since the 4-momenta are all contracted with projection operators introduced in Eq.~\eqref{projector}, we have the freedom to replace all of them by their orthogonal projection with respect to the fluid 4-velocity. Then, 
\begin{eqnarray}
&& \Delta_{\alpha_1 \cdots \alpha_\ell}^{\nu_1 \cdots \nu_\ell}\Delta^{\nu_{\ell+1}}_{\alpha_{\ell+1}}k^{\langle\alpha_{1}\rangle}\cdots k^{\langle\alpha_{\ell+1}\rangle} \notag \\
&=& \Delta_{\alpha_1 \cdots \alpha_\ell}^{\nu_1 \cdots \nu_\ell}\Delta^{\nu_{\ell+1}}_{\alpha_{\ell+1}} \left(k^{\langle\alpha_{1}}\cdots k^{\alpha_{\ell+1}\rangle} -\frac{C(\ell+1,1)}{N_{\ell+1,1}} \sum_{\mathcal{P}^{\ell+1}_\alpha \mathcal{P}^{\ell+1}_\beta} \Delta^{\alpha_\ell \alpha_{\ell+1}}  \Delta_{\beta_\ell \beta_{\ell+1}} \Delta^{\alpha_1}_{\beta_1}\cdots \Delta^{\alpha_{\ell-1}}_{\beta_{\ell-1}} k^{\beta_1}\cdots k^{\beta_{\ell+1}}\right). \label{first_term_termo}
\end{eqnarray}
Here we make use of Eq.~\eqref{projector} to write the irreducible 4-momenta of rank $\ell+1$, where only the first two terms provide non-zero contributions. In fact, all terms involving two or more doubly contravariant (and the same number of doubly covariant) rank 2 projection operators lead to vanishing terms, since there is always a contraction of the type 
\begin{equation}
\Delta_{\alpha_1 \cdots \alpha_\ell}^{\nu_1 \cdots \nu_\ell} \Delta^{\alpha_i \alpha_j} = 0, \ \ \text{with} \ \ 0 \leq (i,j) \leq \ell,
\label{contractions_zero_traceless}
\end{equation}
which is zero, given the traceless property of the projection operator, see Eq.~\eqref{traceless}.

Next, it is necessary to compute the number of non-vanishing permutations accounted in the sums in Eq.~\eqref{first_term_termo}. First, we look at the doubly contravariant projection operator. One of its indices must necessarily be $\alpha_{\ell+1}$, regardless of the other, otherwise leading to zero given Eq.~\eqref{contractions_zero_traceless}. This index can be paired with $\ell$ other indices, which corresponds to a total of $\ell$ combinations. Second, the doubly covariant projection operators are contracted with the corresponding pair of 4-momenta, without restrictions to be imposed on these indices. In this case, the number of different combinations is simply a permutation of $\ell+1$ elements taken in pairs, i.e., $\frac{(\ell+1)!}{2! (\ell-1)!}$. Last, the $\ell-1$ remaining 4-momenta are contracted with projection operators with one covariant and one contravariant index. For a given a covariant (contravariant) index, there are $\ell-1$ possible contravariant (covariant) indices it can be paired with. Naturally, a different covariant (contravariant) index can now be paired with $\ell-2$ contravariant (covariant) indices, and so on. Therefore, the number of permutations for these projectors is simply $(\ell-1)!$, thus leading to
\begin{eqnarray}
k^{\langle\nu_1}\cdots k^{\nu_\ell\rangle}k^{\langle\nu_{\ell+1}\rangle} &=& k^{\langle\nu_{1}}\cdots k^{\nu_{\ell+1}\rangle} - \ell \times \frac{(\ell+1)!}{2! (\ell-1)!} \times (\ell-1)! \times \frac{C(\ell+1,1)}{N_{\ell+1,1}} \Delta_{\alpha_1 \cdots \alpha_\ell}^{\nu_1 \cdots \nu_\ell} \notag \\
&\times& \sum_{\mathcal{P}^{\ell+1}_\alpha \mathcal{P}^{\ell+1}_\beta} \Delta^{\alpha_\ell \alpha_{\ell+1}}  \Delta_{\beta_\ell \beta_{\ell+1}} \Delta^{\alpha_1}_{\beta_1}\cdots \Delta^{\alpha_{\ell-1}}_{\beta_{\ell-1}} k^{\beta_1}\cdots k^{\beta_{\ell+1}} \notag \\
&=& k^{\langle\nu_{1}}\cdots k^{\nu_{\ell+1}\rangle} + b_{\mathbf{k}} \frac{\ell}{2\ell+1} \Delta_{\alpha_1 \cdots \alpha_\ell}^{\nu_1 \cdots \nu_\ell}  \Delta^{\alpha_\ell \nu_{\ell+1}} k^{\langle\alpha_1\rangle}\cdots k^{\langle\alpha_{\ell-1}\rangle} \notag \\
&=& k^{\langle\nu_{1}}\cdots k^{\nu_{\ell+1}\rangle} + b_{\mathbf{k}} \frac{\ell}{2\ell+1} \Delta_{\alpha_1 \cdots \alpha_\ell}^{\nu_1 \cdots \nu_\ell}  \Delta^{\alpha_\ell \nu_{\ell+1}} k^{\langle\alpha_1}\cdots k^{\alpha_{\ell-1}\rangle}, \label{identity1}
\end{eqnarray}
which is exactly Eq.~\eqref{identity2}. Note that we are able to replace $k^{\langle\alpha_1\rangle}\cdots k^{\langle\alpha_{\ell-1}\rangle}$ by $k^{\langle\alpha_1}\cdots k^{\alpha_{\ell-1}\rangle}$ in the last equality since this term is contracted with $\Delta_{\alpha_1 \cdots \alpha_\ell}^{\nu_1 \cdots \nu_\ell}$. Finally, plugging this result in Eq.~\eqref{eq11}, we obtain
\begin{equation}
\Delta^{\mu_1\cdots\mu_\ell}_{\nu_1\cdots\nu_\ell} \left\langle \frac{dE^r_{\mathbf{k}}}{d\tau} k^{\langle\nu_1}\cdots k^{\nu_\ell\rangle} \right\rangle = r \left[\varrho^{\mu_1 \cdots \mu_{\ell+1}}_{r-1} \dot{u}_{\mu_{\ell+1}} + \frac{\ell}{2\ell+1}\left(m^2 \varrho_{r-1}^{\langle\mu_1 \cdots \mu_{\ell-1}} - \varrho_{r+1}^{\langle\mu_1 \cdots \mu_{\ell-1}}\right) \dot{u}^{\mu_\ell\rangle} \right], \label{eq18}
\end{equation}
where we have used $b_{\mathbf{k}} = m^2 - E_{\mathbf{k}}^2$.

\subsection{Second term}

The second term on the right-hand side of Eq.~\eqref{eq2} is
\begin{equation}
\Delta^{\mu_1\cdots\mu_\ell}_{\nu_1\cdots\nu_\ell} \left\langle E^r_{\mathbf{k}} \frac{d}{d\tau}\left(k^{\langle\nu_1}\cdots k^{\nu_\ell\rangle}\right) \right\rangle = \Delta^{\mu_1\cdots\mu_\ell}_{\nu_1\cdots\nu_\ell} \frac{d}{d\tau}\left(\Delta^{\nu_1\cdots\nu_\ell}_{\alpha_1\cdots\alpha_\ell}\right) \left\langle E^r_{\mathbf{k}} k^{\alpha_1}\cdots k^{\alpha_\ell} \right\rangle, \label{second_term_inicio}
\end{equation}
The task here is to calculate the time derivative of the $2\ell$-index projection operator,
\begin{eqnarray}
\Delta^{\mu_1\cdots\mu_\ell}_{\nu_1\cdots\nu_\ell} \frac{d}{d\tau}\left(\Delta^{\nu_1\cdots\nu_\ell}_{\alpha_1\cdots\alpha_\ell}\right)
= 
\Delta_{\nu_1 \cdots \nu_\ell}^{\mu_1 \cdots \mu_\ell} \frac{d}{d\tau} \left( \frac{C(\ell,0)}{N_{\ell,0}} \sum_{\mathcal{P}^\ell_\nu \mathcal{P}^\ell_\alpha} \Delta^{\nu_1}_{\alpha_1} \cdots \Delta^{\nu_\ell}_{\alpha_\ell} + \frac{C(\ell,1)}{N_{\ell,1}} \sum_{\mathcal{P}^\ell_\nu \mathcal{P}^\ell_\alpha} \Delta^{\nu_1 \nu_2} \Delta_{\alpha_1 \alpha_2} \Delta^{\nu_3}_{\alpha_3} \cdots \Delta^{\nu_\ell}_{\alpha_\ell} + \cdots \right).
\end{eqnarray}
It can readily be seen that, except for the first term, there will always be contractions either of the type given by Eq.~\eqref{contractions_zero_traceless}, or
\begin{eqnarray}
\Delta^{\mu_1\cdots\mu_\ell}_{\nu_1\cdots\nu_\ell} u^{\nu_i},  \ \ \text{with} \ \ 0 \leq i \leq \ell,
\label{contractions_zero_orthogonal}
\end{eqnarray}
which are identically zero since the projection operator is traceless and orthogonal to the fluid 4-velocity by construction. Therefore, the only non-vanishing contribution comes from the first term. Analogously to the discussion developed in the previous subsection, the sums account for $\ell!$ possibilities to arrange the indices of the remaining projection operators. In particular, all permutations lead to the same result when contracted with $k^{\alpha_1}\cdots k^{\alpha_\ell}$, this yielding a factor $\ell$. Furthermore, since the covariant derivative of the metric is zero, we have
\begin{equation}
\frac{d}{d\tau} \Delta^\mu_\nu = - \dot{u}^\mu u_\nu - u^\mu \dot{u}_\nu,
\end{equation}
and thus
\begin{equation}
\Delta^{\mu_1\cdots\mu_\ell}_{\nu_1\cdots\nu_\ell} \frac{d}{d\tau}\left(\Delta^{\nu_1\cdots\nu_\ell}_{\alpha_1\cdots\alpha_\ell}\right) = \Delta^{\mu_1\cdots\mu_\ell}_{\nu_1\cdots\nu_\ell}  \frac{C(\ell,0)}{N_{\ell,0}} \frac{d}{d\tau} \sum_{\mathcal{P}^\ell_\nu \mathcal{P}^\ell_\alpha} \Delta^{\nu_1}_{\alpha_1} \cdots \Delta^{\nu_\ell}_{\alpha_\ell} =  -\ell \Delta^{\mu_1\cdots\mu_\ell}_{\nu_1\cdots\nu_\ell} \dot{u}^{\nu_\ell} u_{\alpha_\ell} \Delta^{\nu_1}_{\alpha_1} \cdots \Delta^{\nu_{\ell-1}}_{\alpha_{\ell-1}}.
\end{equation}
Using this result in Eq.~\eqref{second_term_inicio}, we obtain
\begin{eqnarray}
\Delta^{\mu_1\cdots\mu_\ell}_{\nu_1\cdots\nu_\ell} \left\langle E^r_{\mathbf{k}} \frac{d}{d\tau}\left(k^{\langle\nu_1}\cdots k^{\nu_\ell\rangle}\right) \right\rangle &=& - \ell \Delta^{\mu_1\cdots\mu_\ell}_{\nu_1\cdots\nu_\ell} \left\langle E^{r+1}_{\mathbf{k}} k^{\langle\nu_1\rangle} \cdots k^{\langle\nu_\ell\rangle} \right\rangle \dot{u}^{\nu_{\ell-1}} \notag \\
&=& - \ell \Delta^{\mu_1\cdots\mu_\ell}_{\nu_1\cdots\nu_\ell} \left\langle E^{r+1}_{\mathbf{k}} k^{\langle\nu_1} \cdots k^{\nu_\ell\rangle} \right\rangle \dot{u}^{\nu_{\ell-1}} \notag \\
&=& -\ell \varrho^{\langle\mu_1 \cdots \mu_{\ell-1}}_{r+1} \dot{u}^{\mu_\ell\rangle}, \label{eq9}
\end{eqnarray}
where we have used Eq.~\eqref{identity0} to obtain the second equality, whose derivation is discussed in Appendix \ref{app_formulas}. 

\subsection{Third term} 

The last term on the right-hand side of Eq.~\eqref{eq2} is

\begin{equation}
\Delta^{\mu_1\cdots\mu_\ell}_{\nu_1\cdots\nu_\ell}\int dK E^r_{\mathbf{k}} k^{\langle\nu_1}\cdots k^{\nu_\ell\rangle} \frac{d f_{\mathbf{k}}}{d\tau} \label{eq19}
\end{equation}

At this point, it is convenient to decompose the 4-momentum in terms of its longitudinal and transverse components with respect to the fluid 4-velocity as $k^\mu = u^\mu E_{\mathbf{k}} - k^{\langle\mu\rangle}$. Then, from the Boltzmann equation, Eq.~\eqref{boltz}, it is possible to express the time derivative of the single-particle distribution function as
\begin{equation}
\frac{df_{\mathbf{k}}}{d\tau} = E_{\mathbf{k}}^{-1} C[f] - E_{\mathbf{k}}^{-1} k^{\langle\mu\rangle} \nabla_\mu f_{\mathbf{k}}.
\end{equation}
Therefore, Eq.~\eqref{eq19} can be written as
\begin{equation}
\Delta^{\mu_1\cdots\mu_\ell}_{\nu_1\cdots\nu_\ell}\int dK E^r_{\mathbf{k}} k^{\langle\nu_1}\cdots k^{\nu_\ell\rangle} \frac{d f_{\mathbf{k}}}{d\tau} = \Delta^{\mu_1\cdots\mu_\ell}_{\nu_1\cdots\nu_\ell} \left(\int dK E_{\mathbf{k}}^{r-1} k^{\langle\nu_1}\cdots k^{\nu_\ell\rangle} C[f] - \int dK E_{\mathbf{k}}^{r-1} k^{\langle\nu_1}\cdots k^{\nu_\ell\rangle}k^{\langle\mu_{\ell+1}\rangle}\nabla_{\mu_{\ell+1}} f_{\mathbf{k}} \right). \label{eq24}
\end{equation}
The first term on the right-hand side of this equation can be immediately identified as the generalized collision term, see Eq.~\eqref{generalized_collision},
\begin{equation}
\Delta^{\mu_1\cdots\mu_\ell}_{\nu_1\cdots\nu_\ell} \int dK E_{\mathbf{k}}^{r-1} k^{\langle\nu_1}\cdots k^{\nu_\ell\rangle} C[f] = \mathcal{C}^{\langle \mu_1\cdots\mu_\ell \rangle}_{r-1}. \label{eq25}
\end{equation}

We now look at the last term term on the right-hand side of Eq.~\eqref{eq24}.

\begin{eqnarray}
&-& \Delta^{\mu_1\cdots\mu_\ell}_{\nu_1\cdots\nu_\ell} \int dK E_{\mathbf{k}}^{r-1} k^{\langle\nu_1}\cdots k^{\nu_\ell\rangle}k^{\langle\nu_{\ell+1}\rangle}\nabla_{\nu_{\ell+1}} f_{\mathbf{k}} \notag \\
&=& - \Delta^{\mu_1\cdots\mu_\ell}_{\nu_1\cdots\nu_\ell} \left( \nabla_{\nu_{\ell+1}} \left\langle E_{\mathbf{k}}^{r-1} k^{\langle\nu_1}\cdots k^{\nu_\ell\rangle}k^{\langle\nu_{\ell+1}\rangle} \right\rangle \right. + \Delta^{\mu_1\cdots\mu_\ell}_{\nu_1\cdots\nu_\ell} \left\langle \left(\nabla_{\nu_{\ell+1}}E_{\mathbf{k}}^{r-1}\right) k^{\langle\nu_1}\cdots k^{\nu_\ell\rangle}k^{\langle\nu_{\ell+1}\rangle}\right\rangle + \notag \\
&+& \Delta^{\mu_1\cdots\mu_\ell}_{\nu_1\cdots\nu_\ell} \left\langle E_{\mathbf{k}}^{r-1} \nabla_{\nu_{\ell+1}} \left(k^{\langle\nu_1}\cdots k^{\nu_\ell\rangle}k^{\langle\nu_{\ell+1}\rangle}\right) \right\rangle. \label{eq26}
\end{eqnarray}
In the following, we work each of these terms individually.

\subsubsection*{Part I}

Using Eq.~\eqref{identity1}, the first term on the right-hand of Eq.~\eqref{eq26} can be written as
\begin{eqnarray}
&-& \Delta^{\mu_1\cdots\mu_\ell}_{\nu_1\cdots\nu_\ell} \nabla_{\nu_{\ell+1}} \left\langle E_{\mathbf{k}}^{r-1} k^{\langle\nu_1}\cdots k^{\nu_\ell\rangle}k^{\langle\nu_{\ell+1}\rangle} \right\rangle \notag \\
&=& - \Delta^{\mu_1\cdots\mu_\ell}_{\nu_1\cdots\nu_\ell} \nabla_{\nu_{\ell+1}} \left( \left\langle E_{\mathbf{k}}^{r-1} k^{\langle\nu_{1}}\cdots k^{\nu_{\ell+1}\rangle} \right\rangle + \frac{\ell}{2\ell+1} \Delta_{\alpha_1 \cdots \alpha_\ell}^{\nu_1 \cdots \nu_\ell}  \Delta^{\alpha_\ell \nu_{\ell+1}} \left\langle b_{\mathbf{k}} E_{\mathbf{k}}^{r-1} k^{\langle\alpha_1}\cdots k^{\alpha_{\ell-1}\rangle}\right\rangle \right) \notag \\
&=& - \Delta^{\mu_1\cdots\mu_\ell}_{\nu_1\cdots\nu_\ell} \nabla_{\nu_{\ell+1}} \varrho^{\nu_1 \cdots \nu_{\ell+1}}_{r-1} - \frac{\ell}{2\ell+1} \Delta^{\mu_1\cdots\mu_\ell}_{\alpha_1\cdots\alpha_\ell} \nabla^{\alpha_\ell} \left\langle b_{\mathbf{k}} E_{\mathbf{k}}^{r-1} k^{\langle\alpha_1}\cdots k^{\alpha_{\ell-1}\rangle}\right\rangle \notag \\
&=& - \Delta^{\mu_1\cdots\mu_\ell}_{\nu_1\cdots\nu_\ell} \nabla_{\nu_{\ell+1}} \varrho^{\nu_1 \cdots \nu_{\ell+1}}_{r-1} - \frac{\ell}{2\ell+1} \nabla^{\langle\mu_1}\left(m^2 \varrho^{\mu_2\cdots\mu_\ell\rangle}_{r-1} - \varrho^{\mu_2\cdots\mu_\ell\rangle}_{r+1}\right), \label{eq29+33}
\end{eqnarray}
where we have used $\Delta_{\alpha_1 \cdots \alpha_\ell}^{\nu_1 \cdots \nu_\ell} \nabla_{\nu_{\ell+1}} \Delta^{\alpha_\ell \nu_{\ell+1}} = 0$ and $\Delta^{\mu_1 \cdots \mu_\ell}_{\nu_1 \cdots \nu_\ell}\left( \nabla^{\alpha_\ell} \Delta^{\nu_1 \cdots \nu_\ell}_{\alpha_1 \cdots \alpha_\ell}\right) = 0$, since every term carries contractions either of the type given by Eq.~\eqref{contractions_zero_traceless} or by Eq.~\eqref{contractions_zero_orthogonal}.

\subsubsection*{Part II}

We proceed to analyze the second term on the right-hand side of Eq.~\eqref{eq26}. First, we note that
\begin{equation}
\nabla_\mu E_{\mathbf{k}}^{r-1} = (r-1) E_{\mathbf{k}}^{r-2} k^{\langle\nu\rangle} \left( \nabla_\mu u_\nu \right).
\end{equation}
Then, we use the relation
\begin{equation}
\nabla_\mu u_\nu = \sigma_{\mu\nu} +\frac{1}{3}\Delta_{\mu\nu}\theta + \omega_{\mu\nu} \label{eq36}
\end{equation}
to write
\begin{eqnarray}
&& \Delta^{\mu_1\cdots\mu_\ell}_{\nu_1\cdots\nu_\ell} \left\langle \left(\nabla_{\nu_{\ell+1}}E_{\mathbf{k}}^{r-1}\right) k^{\langle\nu_1}\cdots k^{\nu_\ell\rangle}k^{\langle\nu_{\ell+1}\rangle}\right\rangle \notag \\
&=& (r-1) \Delta^{\mu_1\cdots\mu_\ell}_{\nu_1\cdots\nu_\ell} \left(\frac{\theta}{3} \left\langle b_\textbf{k} E_{\mathbf{k}}^{r-2} k^{\langle\nu_1}\cdots k^{\nu_\ell\rangle} \right\rangle+ \left\langle E_{\mathbf{k}}^{r-2} k^{\langle\nu_1}\cdots k^{\nu_\ell\rangle}k^{\langle\nu_{\ell+1}\rangle}k^{\langle\nu_{\ell+2}\rangle} \right\rangle \sigma_{\nu_{\ell+1}\nu_{\ell+2}} \right) \notag \\
&=& \frac{r-1}{3} \left( m^2 \varrho^{\mu_1\cdots\mu_\ell}_{r-2} -  \varrho^{\mu_1\cdots\mu_\ell}_r\right) \theta + (r-1) \Delta^{\mu_1\cdots\mu_\ell}_{\nu_1\cdots\nu_\ell} \left\langle E_{\mathbf{k}}^{r-2} k^{\langle\nu_1}\cdots k^{\nu_\ell\rangle}k^{\langle\nu_{\ell+1}\rangle}k^{\langle\nu_{\ell+2}\rangle} \right\rangle \sigma_{\nu_{\ell+1}\nu_{\ell+2}}  \label{eq40+41}
\end{eqnarray}
The vorticity is contracted with a symmetric tensor, thus leading to zero. In order to express the second term on the right-hand side of Eq.~\eqref{eq40+41} in terms of irreducible moments, we successively use Eq.~\eqref{identity1},
\begin{eqnarray}
&& \Delta^{\mu_1\cdots\mu_\ell}_{\nu_1\cdots\nu_\ell} \left\langle E_{\mathbf{k}}^{r-2} k^{\langle\nu_1}\cdots k^{\nu_\ell\rangle}k^{\langle\nu_{\ell+1}\rangle}k^{\langle\nu_{\ell+2}\rangle} \right\rangle \sigma_{\nu_{\ell+1}\nu_{\ell+2}} \notag \\
&=& \Delta^{\mu_1\cdots\mu_\ell}_{\nu_1\cdots\nu_\ell} \left\langle E_{\mathbf{k}}^{r-2} \left( k^{\langle\nu_{1}}\cdots k^{\nu_{\ell+1}\rangle} k^{\langle\nu_{\ell+2}\rangle} + \frac{\ell}{2\ell+1} b_{\mathbf{k}} \Delta_{\alpha_1 \cdots \alpha_\ell}^{\nu_1 \cdots \nu_\ell}  \Delta^{\alpha_\ell \nu_{\ell+1}} k^{\langle\alpha_1}\cdots k^{\alpha_{\ell-1}\rangle} k^{\langle\nu_{\ell+2}\rangle} \right) \right\rangle  \notag \sigma_{\nu_{\ell+1}\nu_{\ell+2}} \\
&=& \Delta^{\mu_1\cdots\mu_\ell}_{\nu_1\cdots\nu_\ell} \left\langle E_{\mathbf{k}}^{r-2} \left[ k^{\langle\nu_{1}}\cdots k^{\nu_{\ell+2}\rangle} + \frac{\ell+1}{2\ell+3} b_{\mathbf{k}} \Delta^{\nu_1 \cdots \nu_{\ell+1}}_{\beta_1 \cdots \beta_{\ell+1}}\Delta^{\beta_{\ell+1}\nu_{\ell+2}} k^{\langle\beta_1}\cdots k^{\beta_\ell\rangle} + \frac{\ell}{2\ell+1} b_{\mathbf{k}} \Delta_{\alpha_1 \cdots \alpha_\ell}^{\nu_1 \cdots \nu_\ell}  \Delta^{\alpha_\ell \nu_{\ell+1}} \right. \right. \notag \\
&\times& \left. \left. \left( k^{\langle\alpha_1}\cdots k^{\alpha_{\ell-1}} k^{\nu_{\ell+2}\rangle} + \frac{\ell-1}{2\ell-1} b_{\mathbf{k}} \Delta^{\alpha_1 \cdots \alpha_{\ell-1}}_{\lambda_1 \cdots \lambda_{\ell-1}}\Delta^{\lambda_{\ell-1} \nu_{\ell+2}}  k^{\langle\lambda_1}\cdots k^{\lambda_{\ell-2}\rangle} \right) \right] \right\rangle \sigma_{\nu_{\ell+1}\nu_{\ell+2}} \notag \\
&=& \varrho^{\mu_1\cdots \mu_{\ell+2}}_{r-2} \sigma_{\mu_{\ell+1}\mu_{\ell+2}} + \frac{\ell(\ell-1)}{4\ell^2-1} \left( m^4 \varrho^{\langle\mu_1\cdots \mu_{\ell-2}}_{r-2} -2m^2 \varrho^{\langle\mu_1\cdots \mu_{\ell-2}}_{r} + \varrho^{\langle\mu_1\cdots \mu_{\ell-2}}_{r+2} \right) \sigma^{\mu_{\ell-1}\mu_{\ell}\rangle} + \notag \\
&+&\frac{\ell}{2\ell+1} \left( m^2 \varrho^{\alpha \langle \mu_1 \cdots \mu_{\ell-1}}_{r-2} - \varrho^{\alpha \langle \mu_1 \cdots \mu_{\ell-1}}_{r} \right) \sigma^{\mu_\ell\rangle}_\alpha + \frac{\ell+1}{2\ell+3} \Delta^{\mu_1\cdots\mu_\ell}_{\nu_1\cdots\nu_\ell} \Delta^{\nu_1\cdots\nu_{\ell+1}}_{\beta_1\cdots\beta_{\ell+1}} \left( m^2 \varrho^{\beta_1\cdots\beta_\ell}_{r-2} - \varrho^{\beta_1\cdots\beta_\ell}_r  \right) \sigma^{\beta_{\ell+1}}_{\nu_{\ell+1}}. \label{eq45+1}
\end{eqnarray}

Using Eq.~\eqref{projector}, it is possible to write the term $\Delta^{\mu_1\cdots\mu_\ell}_{\nu_1\cdots\nu_\ell} \Delta^{\nu_1\cdots\nu_{\ell+1}}_{\beta_1\cdots\beta_{\ell+1}}\varrho^{\beta_1\cdots\beta_\ell}\sigma^{\beta_{\ell+1}}_{\nu_{\ell+1}}$ as

\begin{eqnarray}
&& \Delta^{\mu_1\cdots\mu_\ell}_{\nu_1\cdots\nu_\ell} \Delta^{\nu_1\cdots\nu_{\ell+1}}_{\beta_1\cdots\beta_{\ell+1}} \varrho^{\beta_1\cdots\beta_\ell}\sigma^{\beta_{\ell+1}}_{\nu_{\ell+1}} \\ 
&=& \Delta^{\mu_1\cdots\mu_\ell}_{\nu_1\cdots\nu_\ell} \left[ \frac{C(\ell+1,0)}{\mathcal{N}_{\ell+1,0}} \sum_{\mathcal{P}^{\ell+1}_\nu \mathcal{P}^{\ell+1}_\beta} \Delta^{\nu_1}_{\beta_1}\cdots \Delta^{\nu_{\ell+1}}_{\beta_{\ell+1}} + \frac{C(\ell+1,1)}{\mathcal{N}_{\ell+1,1}} \sum_{\mathcal{P}^{\ell+1}_\nu \mathcal{P}^{\ell+1}_\beta} \Delta^{\nu_\ell \nu_{\ell+1}} \Delta_{\beta_\ell \beta_{\ell+1}} \Delta^{\nu_1}_{\beta_1}\cdots \Delta^{\nu_{\ell-1}}_{\beta_{\ell-1}} + \cdots \right]\varrho^{\beta_1\cdots\beta_\ell}\sigma^{\beta_{\ell+1}}_{\nu_{\ell+1}}, \notag
\end{eqnarray}
where the ellipsis denote all terms containing at least two covariant (and contravariant) projection operators, which are identically zero since they involve contractions similar to Eq.~\eqref{contractions_zero_traceless}. 

We analyze each term inside square brackets separately, starting with the first. All permutations containing the projector $\Delta^{\nu_{\ell+1}}_{\beta_{\ell+1}}$ are identically zero, since the shear tensor is traceless and orthogonal to the 4-velocity, and thus $\Delta^{\nu_{\ell+1}}_{\beta_{\ell+1}}\sigma^{\beta_{\ell+1}}_{\nu_{\ell+1}}=0$. Therefore, in order to account exclusively the non-vanishing terms, we calculate the number of total permutations and exclude those that contain the aforementioned operator. Hence, the number of contributing terms in the sum is simply
\begin{equation}
(\ell+1)! - \ell!  = \ell \ell!.
\end{equation}
Then, the first term inside square brackets reduces to
\begin{eqnarray}
&& \frac{C(\ell+1,0)}{\mathcal{N}_{\ell+1,0}} \sum_{\mathcal{P}^{\ell+1}_\nu \mathcal{P}^{\ell+1}_\beta} \Delta^{\nu_1}_{\beta_1}\cdots \Delta^{\nu_{\ell+1}}_{\beta_{\ell+1}} \varrho^{\beta_1\cdots\beta_\ell}\sigma^{\beta_{\ell+1}}_{\nu_{\ell+1}} \notag \\
&=& \frac{C(\ell+1,0)}{\mathcal{N}_{\ell+1,0}} \times \ell \ell! \times \Delta^{\nu_1}_{\beta_{\ell+1}} \Delta^{\nu_2}_{\beta_2}\cdots \Delta^{\nu_\ell}_{\beta_\ell}\Delta^{\nu_{\ell+1}}_{\beta_1} \varrho^{\beta_1\cdots\beta_\ell}\sigma^{\beta_{\ell+1}}_{\nu_{\ell+1}} = \frac{\ell}{\ell+1} \Delta^{\nu_1}_{\beta_{\ell+1}} \Delta^{\nu_2}_{\beta_2}\cdots \Delta^{\nu_\ell}_{\beta_\ell}\Delta^{\nu_{\ell+1}}_{\beta_1} \varrho^{\beta_1\cdots\beta_\ell}\sigma^{\beta_{\ell+1}}_{\nu_{\ell+1}} \notag \\
&=& \frac{\ell}{\ell+1} \varrho^{\alpha \nu_1 \cdots \nu_{\ell-1}} \sigma^{\nu_\ell}_\alpha.
\label{eq46+8}
\end{eqnarray}

The next step is to compute the number of non-zero permutations in the second term in square brackets. First, we note that the doubly contravariant projection operator must be  $\Delta^{\nu_i \nu_{\ell+1}}$, with $0 \leq i \leq \ell$, which accounts for $\ell$ different possibilities, otherwise leading to vanishing terms analogous to Eq.~\eqref{contractions_zero_traceless}. Similarly, the doubly covariant projection operator can only be of the type $\Delta_{\beta_i \beta_{\ell+1}}$, once again corresponding to $\ell$ different combinations. In particular, since the indices $\beta_{\ell+1}$ and $\nu_{\ell+1}$ were already used in constructing these projectors, there are no restrictions to be imposed on the projectors with one covariant and one contravariant indices. Therefore, the first contravariant (covariant) index can be paired with any of the remaining $\ell-1$ covariant (contravariant) indices. The next can be paired with $\ell-2$ indices and so on. Wherefore, the number of permutations in this case is $(\ell-1)\times (\ell-2) \times \cdots \times 1=(\ell-1)!$, and thus
\begin{eqnarray}
&& \frac{C(\ell+1,1)}{\mathcal{N}_{\ell+1,1}} \sum_{\mathcal{P}^{\ell+1}_\nu \mathcal{P}^{\ell+1}_\beta} \Delta^{\nu_\ell \nu_{\ell+1}} \Delta_{\beta_\ell \beta_{\ell+1}} \Delta^{\nu_1}_{\beta_1}\cdots \Delta^{\nu_{\ell-1}}_{\beta_{\ell-1}} \varrho^{\beta_1\cdots\beta_\ell}\sigma^{\beta_{\ell+1}}_{\nu_{\ell+1}} \notag \\
&=& \frac{C(\ell+1,1)}{\mathcal{N}_{\ell+1,1}} \times \ell \times \ell \times (\ell-1)! \times \Delta^{\nu_\ell \nu_{\ell+1}} \Delta_{\beta_\ell \beta_{\ell+1}} \Delta^{\nu_1}_{\beta_1}\cdots \Delta^{\nu_{\ell-1}}_{\beta_{\ell-1}} \varrho^{\beta_1\cdots\beta_\ell}\sigma^{\beta_{\ell+1}}_{\nu_{\ell+1}} \notag \\
&=& -\frac{2\ell}{(\ell+1)(2\ell+1)} \varrho^{\alpha \nu_1 \cdots \nu_{\ell-1}} \sigma^{\nu_\ell}_\alpha.
\label{eq46+4}
\end{eqnarray}
Then, from Eqs.~\eqref{eq46+8} and \eqref{eq46+4}, the last term on the right-hand side of Eq.~\eqref{eq45+1} becomes

\begin{equation}
\frac{\ell+1}{2\ell+3} \Delta^{\mu_1\cdots\mu_\ell}_{\nu_1\cdots\nu_\ell} \Delta^{\nu_1\cdots\nu_{\ell+1}}_{\beta_1\cdots\beta_{\ell+1}} \left( m^2 \varrho^{\beta_1\cdots\beta_\ell}_{r-2} - \varrho^{\beta_1\cdots\beta_\ell}_r  \right) \sigma^{\beta_{\ell+1}}_{\nu_{\ell+1}} = \frac{\ell(2\ell-1)}{(2\ell+1)(2\ell+3)} \left( m^2 \varrho^{\alpha\langle\mu_1\cdots\mu_{\ell-1}}_{r-2} - \varrho^{\alpha\langle\mu_1\cdots\mu_{\ell-1}}_r  \right) \sigma^{\mu_{\ell}\rangle}_\alpha. \label{eq46+13}
\end{equation}

Finally, from Eqs.~\eqref{eq40+41}, \eqref{eq45+1} and \eqref{eq46+13}, the second term on the right-hand side of Eq.~\eqref{eq26} is given by 

\begin{eqnarray}
&& (r-1) \Delta^{\mu_1\cdots\mu_\ell}_{\nu_1\cdots\nu_\ell} \left\langle \left(\nabla_{\nu_{\ell+1}}E_{\mathbf{k}}^{r-1}\right) k^{\langle\nu_1}\cdots k^{\nu_\ell\rangle}k^{\langle\nu_{\ell+1}\rangle}\right\rangle \notag \\
&=& (r-1) \varrho^{\mu_1\cdots \mu_{\ell+2}}_{r-2} \sigma_{\mu_{\ell+1}\mu_{\ell+2}} + (r-1) \frac{\ell(\ell-1)}{4\ell^2-1} \left( m^4 \varrho^{\langle\mu_1\cdots \mu_{\ell-2}}_{r-2} -2m^2 \varrho^{\langle\mu_1\cdots \mu_{\ell-2}}_{r} + \varrho^{\langle\mu_1\cdots \mu_{\ell-2}}_{r+2} \right) \sigma^{\mu_{\ell-1}\mu_{\ell}\rangle} + \notag \\
&+& (r-1) \frac{2\ell}{2\ell+3} \left( m^2 \varrho^{\alpha\langle\mu_1\cdots\mu_{\ell-1}}_{r-2} - \varrho^{\alpha\langle\mu_1\cdots\mu_{\ell-1}}_r  \right) \sigma^{\mu_{\ell}\rangle}_\alpha + \frac{r-1}{3} \left( m^2 \varrho^{\mu_1\cdots\mu_\ell}_{r-2} -  \varrho^{\mu_1\cdots\mu_\ell}_r\right) \theta. \label{eq46}
\end{eqnarray}

\subsubsection*{Part III}

The final step is to compute the last term on the right-hand side of Eq.~\eqref{eq26}. We have
\begin{eqnarray}
\Delta^{\mu_1\cdots\mu_\ell}_{\nu_1\cdots\nu_\ell} \left\langle E_{\mathbf{k}}^{r-1} \nabla_{\nu_{\ell+1}} \left(k^{\langle\nu_1}\cdots k^{\nu_\ell\rangle}k^{\langle\nu_{\ell+1}\rangle}\right) \right\rangle
= 
\Delta^{\mu_1\cdots\mu_\ell}_{\nu_1\cdots\nu_\ell} \nabla_{\nu_{\ell+1}} \left( \Delta^{\nu_1\cdots\nu_\ell}_{\alpha_1\cdots\alpha_\ell} \Delta^{\nu_{\ell+1}}_{\alpha_{\ell+1}} \right)   \left\langle E_{\mathbf{k}}^{r-1} k^{\alpha_1}\cdots k^{\alpha_{\ell+1}} \right\rangle, \label{eq47}
\end{eqnarray}
where

\begin{equation}
\nabla_{\nu_{\ell+1}} \left( \Delta^{\nu_1\cdots\nu_\ell}_{\alpha_1\cdots\alpha_\ell} \Delta^{\nu_{\ell+1}}_{\alpha_{\ell+1}} \right) = \nabla_{\alpha_{\ell+1}} \Delta^{\nu_1\cdots\nu_\ell}_{\alpha_1\cdots\alpha_\ell} - \Delta^{\nu_1\cdots\nu_\ell}_{\alpha_1\cdots\alpha_\ell} u_{\alpha_{\ell+1}} \theta.
\end{equation}
Therefore, Eq.~\eqref{eq47} becomes

\begin{eqnarray}
&& \Delta^{\mu_1\cdots\mu_\ell}_{\nu_1\cdots\nu_\ell} \nabla_{\nu_{\ell+1}} \left( \Delta^{\nu_1\cdots\nu_\ell}_{\alpha_1\cdots\alpha_\ell} \Delta^{\nu_{\ell+1}}_{\alpha_{\ell+1}} \right)   \left\langle E_{\mathbf{k}}^{r-1} k^{\alpha_1}\cdots k^{\alpha_{\ell+1}} \right\rangle \notag \\
&=& \Delta^{\mu_1\cdots\mu_\ell}_{\nu_1\cdots\nu_\ell}  \left( \nabla_{\alpha_{\ell+1}} \Delta^{\nu_1\cdots\nu_\ell}_{\alpha_1\cdots\alpha_\ell} \right)   \left\langle E_{\mathbf{k}}^{r-1} k^{\alpha_1}\cdots k^{\alpha_{\ell+1}} \right\rangle - \Delta^{\mu_1\cdots\mu_\ell}_{\alpha_1\cdots\alpha_\ell} \left\langle E_{\mathbf{k}}^{r} k^{\alpha_1}\cdots k^{\alpha_{\ell}} \right\rangle \theta \notag \\
&=& \Delta^{\mu_1\cdots\mu_\ell}_{\nu_1\cdots\nu_\ell}  \left( \nabla_{\alpha_{\ell+1}} \Delta^{\nu_1\cdots\nu_\ell}_{\alpha_1\cdots\alpha_\ell} \right)   \left\langle E_{\mathbf{k}}^{r-1} k^{\alpha_1}\cdots k^{\alpha_{\ell+1}} \right\rangle - \varrho^{\mu_1 \cdots \mu_\ell}_r\theta. \label{eq51}
\end{eqnarray}

The next step is to compute the term $\Delta^{\mu_1\cdots\mu_\ell}_{\nu_1\cdots\nu_\ell}  \left( \nabla_{\alpha_{\ell+1}} \Delta^{\nu_1\cdots\nu_\ell}_{\alpha_1\cdots\alpha_\ell} \right)$,
\begin{eqnarray}
\Delta^{\mu_1\cdots\mu_\ell}_{\nu_1\cdots\nu_\ell}  \left( \nabla_{\alpha_{\ell+1}} \Delta^{\nu_1\cdots\nu_\ell}_{\alpha_1\cdots\alpha_\ell} \right) = \Delta^{\mu_1\cdots\mu_\ell}_{\nu_1\cdots\nu_\ell} \nabla_{\alpha_{\ell+1}} \left[ \frac{C(\ell,0)}{\mathcal{N}_{\ell,0}} \sum_{\mathcal{P}^{\ell}_\nu \mathcal{P}^{\ell}_\alpha} \Delta^{\nu_1}_{\alpha_1}\cdots \Delta^{\nu_{\ell}}_{\alpha_{\ell}} + \cdots \right].
\end{eqnarray}
Once again, the ellipsis denote all vanishing terms. In fact, terms containing at least one doubly covariant (and one corresponding doubly contravariant) projection operators are identically zero, since they always involve contractions analogous to Eqs.~\eqref{contractions_zero_traceless} and/or \eqref{contractions_zero_orthogonal}. Furthermore, given the symmetry of the projection operator being contracted, all terms yield the same result, 
\begin{equation}
\ell \Delta^{\mu_1\cdots\mu_\ell}_{\nu_1\cdots\nu_\ell} \frac{C(\ell,0)}{\mathcal{N}_{\ell,0}} \sum_{\mathcal{P}^{\ell}_\nu \mathcal{P}^{\ell}_\alpha} \Delta^{\nu_1}_{\alpha_1}\cdots \Delta^{\nu_{\ell-1}}_{\alpha_{\ell-1}} \left( \nabla_{\alpha_{\ell+1}} \Delta^{\nu_{\ell}}_{\alpha_{\ell}} \right) ,
\end{equation}
where $\nabla_{\alpha_{\ell+1}} \Delta^{\nu_i}_{\alpha_j} = - \left(u^{\nu_i} \nabla_{\alpha_{\ell+1}} u_{\alpha_j} + u_{\alpha_j} \nabla_{\alpha_{\ell+1}} u^{\nu_i} \right)$, with $\left\{ i, j \right\} \in \left[ 1, \dots, \ell\right]$. 
The $\ell-1$ remaining projection operators that are not being derived are contracted with $\Delta^{\mu_1\cdots\mu_\ell}_{\nu_1\cdots\nu_\ell}$. We remark that this is a symmetric tensor under the exchange of its covariant and contravariant indices, and therefore all these terms are identical, hence the factor $\ell$, which comes from the product rule. As it was already discussed, the number of different ways we can arrange the indices of the projectors $\Delta^{\nu_i}_{\alpha_j}$ is simply $\ell!$, thus leading to
\begin{equation}
\ell \Delta^{\mu_1\cdots\mu_\ell}_{\nu_1\cdots\nu_\ell} \frac{C(\ell,0)}{\mathcal{N}_{\ell,0}} \sum_{\mathcal{P}^{\ell}_\nu \mathcal{P}^{\ell}_\alpha} \Delta^{\nu_1}_{\alpha_1}\cdots \Delta^{\nu_{\ell-1}}_{\alpha_{\ell-1}} \left( \nabla_{\alpha_{\ell+1}} \Delta^{\nu_{\ell}}_{\alpha_{\ell}} \right) 
= - \ell \Delta^{\mu_1\cdots\mu_\ell}_{\nu_1\cdots\nu_\ell} \Delta^{\nu_1}_{\alpha_1}\cdots \Delta^{\nu_{\ell-1}}_{\alpha_{\ell-1}} u_{\alpha_\ell} \left( \sigma_{\alpha_{\ell+1}}^{\nu_{\ell}} + \frac{1}{3} \Delta_{\alpha_{\ell+1}}^{\nu_{\ell}}\theta + \omega_{\alpha_{\ell+1}}{}^{\nu_{\ell}} \right), \label{eq57}
\end{equation}
where we have used Eq.~\eqref{eq36} to obtain the last equality. Using this result in Eq.~\eqref{eq51}, 

\begin{eqnarray}
&& \Delta^{\mu_1\cdots\mu_\ell}_{\nu_1\cdots\nu_\ell}  \left( \nabla_{\alpha_{\ell+1}} \Delta^{\nu_1\cdots\nu_\ell}_{\alpha_1\cdots\alpha_\ell} \right)   \left\langle E_{\mathbf{k}}^{r-1} k^{\alpha_1}\cdots k^{\alpha_{\ell+1}} \right\rangle \notag \\
&=& - \ell \Delta^{\mu_1\cdots\mu_\ell}_{\nu_1\cdots\nu_\ell} \Delta^{\nu_1}_{\alpha_1}\cdots \Delta^{\nu_{\ell-1}}_{\alpha_{\ell-1}} u_{\alpha_\ell} \left\langle E_{\mathbf{k}}^{r-1} k^{\alpha_1}\cdots k^{\alpha_{\ell+1}} \right\rangle \left( \sigma_{\alpha_{\ell+1}}^{\nu_{\ell}} + \frac{1}{3} \Delta_{\alpha_{\ell+1}}^{\nu_{\ell}}\theta + \omega_{\alpha_{\ell+1}}{}^{\nu_{\ell}} \right) - \varrho^{\mu_1 \cdots \mu_\ell}_r\theta \notag \\
&=& - \ell \Delta^{\mu_1\cdots\mu_\ell}_{\nu_1\cdots\nu_\ell} \left\langle E_{\mathbf{k}}^{r} k^{\langle \nu_1 \rangle}\cdots k^{\langle \nu_{\ell-1} \rangle} k^{\langle \nu_{\ell+1} \rangle}  \right\rangle \left( \sigma_{\nu_{\ell+1}}^{\nu_{\ell}} + \frac{1}{3} \Delta_{\nu_{\ell+1}}^{\nu_{\ell}}\theta + \omega_{\nu_{\ell+1}}{}^{\nu_{\ell}} \right) - \varrho^{\mu_1 \cdots \mu_\ell}_r\theta \notag \\
&=& - \frac{\ell}{3} \Delta^{\mu_1\cdots\mu_\ell}_{\nu_1\cdots\nu_\ell} \left\langle E_{\mathbf{k}}^{r} k^{\langle \nu_1 \rangle}\cdots k^{\langle \nu_{\ell} \rangle}  \right\rangle\theta - \ell \Delta^{\mu_1\cdots\mu_\ell}_{\nu_1\cdots\nu_\ell} \left\langle E_{\mathbf{k}}^{r} k^{\langle \nu_1 \rangle}\cdots k^{\langle \nu_{\ell-1} \rangle} k^{\langle \nu_{\ell+1} \rangle}  \right\rangle \left( \sigma_{\nu_{\ell+1}}^{\nu_{\ell}} + \omega_{\nu_{\ell+1}}{}^{\nu_{\ell}} \right) - \varrho^{\mu_1 \cdots \mu_\ell}_r\theta \notag \\
&=& - \left( 1 + \frac{\ell}{3} \right) \varrho^{\mu_1\cdots \mu_{\ell}}_r \theta - \ell \Delta^{\mu_1\cdots\mu_\ell}_{\nu_1\cdots\nu_\ell} \left\langle E_{\mathbf{k}}^{r} k^{\langle \nu_1 \rangle}\cdots k^{\langle \nu_{\ell-1} \rangle} k^{\langle \nu_{\ell+1} \rangle}  \right\rangle \left( \sigma_{\nu_{\ell+1}}^{\nu_{\ell}} + \omega_{\nu_{\ell+1}}{}^{\nu_{\ell}} \right). \label{eq58+60}
\end{eqnarray}
The final step is to determine the second term on the right-hand side of Eq.~\eqref{eq58+60}. At this point, we make use of Eq.~\eqref{identity0} to write

\begin{equation}
\Delta^{\mu_1\cdots\mu_\ell}_{\nu_1\cdots\nu_\ell} k^{\langle \nu_1 \rangle}\cdots k^{\langle \nu_{\ell-1} \rangle} k^{\langle \nu_{\ell+1} \rangle} 
= \Delta^{\mu_1\cdots\mu_\ell}_{\nu_1\cdots\nu_\ell} k^{\langle \nu_1}\cdots k^{\nu_{\ell-1}} k^{\nu_{\ell+1} \rangle} + \frac{\ell-1}{2\ell-1} b_{\mathbf{k}} \Delta^{\nu_{\ell-1} \nu_{\ell+1}} k^{\langle \nu_1\rangle} \cdots k^{\langle \nu_{\ell-2}\rangle} + \cdots. \label{eq61}
\end{equation}
As usual, the ellipsis contain all the terms that are zero when contracted with $\Delta^{\mu_1\cdots\mu_\ell}_{\nu_1\cdots\nu_\ell}$. In particular, we note that the doubly contravariant projector must be $\Delta^{\nu_i \nu_{\ell+2}}$, with $i \in \left[1, \dots, \ell\right]$, corresponding to $\ell-1$ different permutations, otherwise leading to contractions analogous to Eq.~\eqref{contractions_zero_traceless}. Therefore, the last term on the right-hand side of Eq.~\eqref{eq58+60} can be cast in the following form 
\begin{eqnarray}
&& - \ell \Delta^{\mu_1\cdots\mu_\ell}_{\nu_1\cdots\nu_\ell} \left\langle E_{\mathbf{k}}^{r} k^{\langle \nu_1 \rangle}\cdots k^{\langle \nu_{\ell-1} \rangle} k^{\langle \nu_{\ell+1} \rangle}  \right\rangle \left( \sigma_{\nu_{\ell+1}}^{\nu_{\ell}} + \omega_{\nu_{\ell+1}}{}^{\nu_{\ell}} \right) \notag \\
&=& - \Delta^{\mu_1\cdots\mu_\ell}_{\nu_1\cdots\nu_\ell} \left( \ell \left\langle E_{\mathbf{k}}^{r} k^{\langle \nu_1}\cdots k^{\nu_{\ell-1}} k^{\nu_{\ell+1} \rangle}  \right\rangle + \frac{\ell(\ell-1)}{2\ell-1} \Delta^{\nu_{\ell-1} \nu_{\ell+1}} \left\langle E_{\mathbf{k}}^{r} b_{\mathbf{k}} k^{\langle \nu_1\rangle} \cdots k^{\langle \nu_{\ell-2}\rangle} \right\rangle \right) \left( \sigma_{\nu_{\ell+1}}^{\nu_{\ell}} + \omega_{\nu_{\ell+1}}{}^{\nu_{\ell}} \right) \notag \\
&=& - \ell \Delta^{\mu_1\cdots\mu_\ell}_{\nu_1\cdots\nu_\ell} \varrho^{ \nu_1 \cdots \nu_{\ell-1}\nu_{\ell+1}}_r \left( \sigma_{\nu_{\ell+1}}^{\nu_{\ell}} - \omega^{\nu_{\ell}}{}_{\nu_{\ell+1}} \right) - \frac{\ell(\ell-1)}{2\ell-1} \Delta^{\mu_1\cdots\mu_\ell}_{\nu_1\cdots\nu_\ell} \left( m^2 \varrho^{\nu_1 \cdots \nu_{\ell-2}}_r - \varrho^{\nu_1 \cdots \nu_{\ell-2}}_{r+2} \right) \left( \sigma^{\nu_{\ell-1}\nu_{\ell}} - \omega^{\nu_{\ell-1}\nu_{\ell}} \right) \notag \\
&=& - \ell \varrho^{\alpha \langle \mu_1 \cdots \mu_{\ell-1}}_r \left( \sigma_\alpha^{\mu_{\ell}\rangle} - \omega^{\mu_{\ell}\rangle}{}_\alpha \right) - \frac{\ell(\ell-1)}{2\ell-1} \left( m^2 \varrho^{\langle \mu_1 \cdots \mu_{\ell-2}}_r - \varrho^{\langle \mu_1 \cdots \mu_{\ell-2}}_{r+2} \right) \sigma^{\mu_{\ell-1} \mu_\ell\rangle}. \label{eq62}
\end{eqnarray}
Note that we have exchanged the order of the covariant and contravariant indices in the vorticity tensor, thus leading to a minus sign, since this is an antisymmetric object, which further leads to $\Delta^{\mu_1\cdots\mu_\ell}_{\nu_1\cdots\nu_\ell} \omega^{\nu_{\ell-1}\nu_{\ell}} = 0$, since the projection operator is symmetric by construction. Then, using Eq.~\eqref{eq62}, we can write Eq.~\eqref{eq58+60} as 

\begin{eqnarray}
&& \Delta^{\mu_1\cdots\mu_\ell}_{\nu_1\cdots\nu_\ell}  \left( \nabla_{\alpha_{\ell+1}} \Delta^{\nu_1\cdots\nu_\ell}_{\alpha_1\cdots\alpha_\ell} \right)   \left\langle E_{\mathbf{k}}^{r-1} k^{\alpha_1}\cdots k^{\alpha_{\ell+1}} \right\rangle \notag \\
&=& -\left(1+\frac{\ell}{3} \right)\varrho^{\mu_1\cdots \mu_{\ell}}_r \theta - \ell \varrho^{\alpha \langle \mu_1 \cdots \mu_{\ell-1}}_r \left( \sigma_\alpha^{\mu_{\ell}\rangle} - \omega^{\mu_{\ell}\rangle}{}_\alpha \right) - \frac{\ell(\ell-1)}{2\ell-1} \left( m^2 \varrho^{\langle \mu_1 \cdots \mu_{\ell-2}}_r - \varrho^{\langle \mu_1 \cdots \mu_{\ell-2}}_{r+2} \right) \sigma^{\mu_{\ell-1} \mu_\ell\rangle}. \label{eq63}
\end{eqnarray}

From Eqs.~\eqref{eq29+33}, \eqref{eq46}, and \eqref{eq63}, it is possible to write Eq.~\eqref{eq26} as 

\begin{eqnarray}
&-& \Delta^{\mu_1\cdots\mu_\ell}_{\nu_1\cdots\nu_\ell} \int dK E_{\mathbf{k}}^{r-1} k^{\langle\nu_1}\cdots k^{\nu_\ell\rangle}k^{\langle\nu_{\ell+1}\rangle}\nabla_{\nu_{\ell+1}} f_{\mathbf{k}} \notag \\
&=& (r-1) \varrho^{\mu_1\cdots \mu_{\ell+2}}_{r-2} \sigma_{\mu_{\ell+1}\mu_{\ell+2}} + \frac{\ell(\ell-1)}{4\ell^2-1} \left[ m^4 \varrho^{\langle\mu_1\cdots \mu_{\ell-2}}_{r-2} - (2r+2\ell-1)m^2 \varrho^{\langle\mu_1\cdots \mu_{\ell-2}}_{r} + (r+2\ell)\varrho^{\langle\mu_1\cdots \mu_{\ell-2}}_{r+2} \right] \sigma^{\mu_{\ell-1}\mu_{\ell}\rangle} + \notag \\
&+& \frac{\ell}{2\ell+3} \left[ (2r-2) m^2 \varrho^{\alpha\langle\mu_1\cdots\mu_{\ell-1}}_{r-2} - (2r+2\ell+1) \varrho^{\alpha\langle\mu_1\cdots\mu_{\ell-1}}_r  \right] \sigma^{\mu_{\ell}\rangle}_\alpha - \left(1+\frac{\ell}{3} \right)\varrho^{\mu_1\cdots \mu_{\ell}}_r \theta + \ell \varrho^{\alpha \langle \mu_1 \cdots \mu_{\ell-1}}_r  \omega^{\mu_{\ell}\rangle}{}_\alpha + \notag \\
&-& \frac{\ell}{2\ell+1} \nabla^{\langle\mu_1}\left(m^2 \varrho^{\mu_2\cdots\mu_\ell\rangle}_{r-1} - \varrho^{\mu_2\cdots\mu_\ell\rangle}_{r+1}\right) - \Delta^{\mu_1\cdots\mu_\ell}_{\nu_1\cdots\nu_\ell} \nabla_{\nu_{\ell+1}} \varrho^{\nu_1 \cdots \nu_{\ell+1}}_{r-1}. 
\label{eq_inf}
\end{eqnarray}

We are finally in position to obtain the final form of the equation of motion for the irreducible moments of rank $\ell$. Combining Eqs.~\eqref{eq18}, \eqref{eq9}, \eqref{eq25} and \eqref{eq_inf}, it is possible to write Eq.~\eqref{eq2} as 

\begin{eqnarray}
\dot{\varrho}^{\langle\mu_1\cdots\mu_\ell\rangle}_r&=&\mathcal{C}^{\mu_1\cdots\mu_\ell}_{r-1}+r\varrho^{\mu_1\cdots\mu_{\ell+1}}_{r-1}\dot{u}_{\mu_{\ell+1}}-\Delta^{\mu_1\cdots\mu_\ell}_{\nu_1\cdots\nu_\ell}\nabla_{\nu_{\ell+1}}\varrho^{\nu_1\cdots\nu_{\ell+1}}_{r-1}+(r-1)\varrho^{\mu_1\cdots\mu_{\ell+2}}_{r-2}\sigma_{\mu_{\ell+1}\mu_{\ell+2}}+\ell\varrho^{\alpha\langle\mu_1\cdots\mu_{\ell-1}}_r\omega^{\mu_\ell\rangle}{}_{\alpha}+\notag \\
&+&\frac{\ell}{2\ell+1}\left[rm^2\varrho_{r-1}^{\langle\mu_1\cdots\mu_{\ell-1}}-(r+2\ell+1)\varrho_{r+1}^{\langle\mu_1\cdots\mu_{\ell-1}}\right]\dot{u}^{\mu_\ell\rangle}+\frac{1}{3}\left[(r-1)m^2\varrho_{r-2}^{\mu_1\cdots\mu_\ell}-(r+\ell+2)\varrho_{r}^{\mu_1\cdots\mu_\ell}\right]\theta+\notag\\
&+&\frac{\ell}{2\ell+3}\left[(2r-2)m^2\varrho_{r-2}^{\alpha\langle\mu_1\cdots\mu_{\ell-1}}-(2r+2\ell+1)\varrho_{r}^{\alpha\langle\mu_1\cdots\mu_{\ell-1}}\right]\sigma_\alpha^{\mu_\ell\rangle}-\frac{\ell}{2\ell+1}\nabla^{\langle\mu_1}\left(m^2\varrho_{r-1}^{\mu_2\cdots\mu_{\ell}\rangle}-\varrho_{r+1}^{\mu_2\cdots\mu_{\ell}\rangle}\right)+\notag\\
&+&\frac{\ell(\ell-1)}{4\ell^2-1}\left[(r-1)m^4\varrho_{r-2}^{\langle\mu_1\cdots\mu_{\ell-2}}-(2r+2\ell-1)m^2\varrho_{r}^{\langle\mu_1\cdots\mu_{\ell-2}}+(r+2\ell)\varrho_{r+2}^{\langle\mu_1\cdots\mu_{\ell-2}}\right]\sigma^{\mu_{\ell-1}\mu_\ell\rangle},
\end{eqnarray}
which is exactly Eq.~\eqref{eq:eom_moms} from the main text.

\section{General identity of the irreducible 4-momenta}
\label{app_formulas}

In this appendix, we derive an explicit relation for the irreducible decomposition of 4-momenta,
\begin{eqnarray}
k^{\langle\mu_1}\cdots k^{\mu_\ell\rangle} & = & \Delta^{\mu_1 \cdots \mu_\ell}_{\nu_1 \cdots \nu_\ell} k^{\nu_1}\cdots k^{\nu_\ell} \notag \\
& = & \left [\sum_{q=0}^{\left[\ell/2\right]}\frac{C(\ell,q)}{\mathcal{N}_{\ell,q}} \sum_{\mathcal{P}^\ell_\mu\mathcal{P}^\ell_\nu}\Delta^{\mu_1\mu_2} \cdots \Delta^{\mu_{2q-1}\mu_{2q}} \Delta_{\nu_1\nu_2} \cdots \Delta_{\nu_{2q-1}\nu_{2q}} \Delta^{\mu_{2q+1}}_{\nu_{2q+1}} \cdots \Delta^{\mu_\ell}_{\nu_\ell} \right] k^{\nu_1}\cdots k^{\nu_\ell}.
\end{eqnarray}
We must analyze all different contractions between the $2-$index projection operators and the 4-momenta. First, we note that the doubly covariant projectors, $\Delta_{\nu_i \nu_j}$, are always contracted with a pair of 4-momenta $k^{\nu_i} k^{\nu_j}$, resulting in $\Delta_{\alpha \beta} k^\alpha k^\beta = b_{\mathbf{k}}$, where $i, j = 1, 2, \dots, 2q$. Naturally, since $q$ corresponds to the number of doubly covariant (as well as the number of doubly contravariant) projection operators, it will also indicate the power of $b_{\mathbf{k}}$ of each term of the sum. It is then necessary to compute the number of different configurations among the indices of aforementioned projectors. The result is simply
\begin{eqnarray}
\sum_{\mathcal{P}^\ell_\nu} \Delta_{\nu_1 \nu_2} \Delta_{\nu_3 \nu_4} \cdots \Delta_{\nu_{2q-1} \nu_{2q}} = \frac{\ell!}{2! (\ell-2)!} \times \frac{(\ell-2)!}{2! (\ell-4)!} \times \cdots \times \frac{(\ell-2q+2)!}{2! (\ell-2q)!} = \frac{\ell!}{2^q (\ell-2q)!} \Delta_{\nu_1 \nu_2} \Delta_{\nu_3 \nu_4} \cdots \Delta_{\nu_{2q-1} \nu_{2q}}.
\end{eqnarray}
Furthermore, these projectors commute. Therefore, in order to avoid counting the same term more than once, we must then multiply this result by the inverse number of different forms that these projectors can be arranged, which is given simply by $1/q!$. 

On the other hand, projectors of the type $\Delta_{\nu_n}^{\mu_m}$ are contracted with
the remaining 4-momenta, $k^{\nu_n}$, with $m, n = 2q+1, 2q+2, \cdots, \ell$. Therefore, each covariant (contravariant) index $\mu_m$ can be paired with a total of $\ell-2q$ contravariant (covariant) indices $\nu_n$, thus resulting in a total of $\ell-2q$ possibilities. In the next projector, the covariant (contravariant) index can now be paired with one of the $\ell-2q-1$ remaining contravariant (covariant) indices, and so on. Consequently, the number of possible configurations to arrange these projectors is $(\ell-2q)!$. 

Then, the irreducible momenta can be expressed as
\begin{equation}
k^{\langle\mu_1}\cdots k^{\mu_\ell\rangle} = \sum_{q=0}^{\left[\ell/2\right]}\frac{C(\ell,q)}{\mathcal{N}_{\ell,q}} \sum_{\mathcal{P}^\ell_\mu} b_{\mathbf{k}}^q \times \frac{\ell!}{2^q (\ell-2q)!} \times \frac{1}{q!} \times (\ell-2q)! \times k^{\langle \mu_{2q+1} \rangle} \cdots k^{\langle \mu_\ell \rangle}.
\end{equation}
Finally, we obtain
\begin{equation}
k^{\langle\mu_1}\cdots k^{\mu_\ell\rangle} = k^{\langle\mu_1\rangle}\cdots k^{\langle\mu_\ell\rangle} + \sum_{q=1}^{[\ell/2]}b_\mathbf{k}^q\frac{\ell!}{2^q q!}\frac{C(\ell,q)}{\mathcal{N}_{\ell,q}}\sum_{\mathcal{P}^\ell_\mu}\Delta^{\mu_1\mu_2}\cdots\Delta^{\mu_{2q-1}\mu_{2q}}k^{\langle\mu_{2q+1}\rangle}\cdots k^{\langle\mu_\ell\rangle},
\end{equation}
which is exactly Eq.~\eqref{identity0} from the main text. 

\bibliographystyle{apsrev4-1}
\bibliography{refs}

\end{document}